\documentclass{article}

\PassOptionsToPackage{numbers, compress}{natbib}
\usepackage[preprint]{neurips_2026}

\usepackage[utf8]{inputenc} 
\usepackage[T1]{fontenc}    
\usepackage{hyperref}       
\usepackage{url}            
\usepackage{booktabs}       
\usepackage{amsfonts}       
\usepackage{nicefrac}       
\usepackage{microtype}      
\usepackage{xcolor}         
\usepackage{graphicx}
\usepackage{enumitem}
\usepackage{wrapfig}
\usepackage{subcaption}
\usepackage{titlesec}
\usepackage{xcolor}
\usepackage{tcolorbox}
\usepackage{multirow}
\tcbuselibrary{breakable}
\titlespacing{\section}{0pt}{1.6ex plus 0.4ex minus 0.2ex}{1.0ex plus 0.2ex}
\titlespacing{\subsection}{0pt}{1.2ex plus 0.3ex minus 0.2ex}{0.6ex plus 0.2ex}
\titlespacing{\subsubsection}{0pt}{1.0ex plus 0.3ex minus 0.2ex}{0.4ex plus 0.2ex}
\titlespacing{\paragraph}{0pt}{0.6ex plus 0.2ex minus 0.1ex}{0.8em}

\title{GazeMind: A Gaze-Guided LLM Agent for Personalized Cognitive Load Assessment}

\author{
  \textbf{Bin Wang$^{1,2}$} \thanks{This work is done during Bin Wang's internship in Meta Reality Labs Research at Redmond, WA.} \quad 
  \textbf{Yue Liu$^{1}$} \quad
  \textbf{Benjamin Newman$^{1}$} \quad
  \textbf{Ajoy S.\ Fernandes$^{1}$} \\
  \textbf{Zhiyuan Wang$^{1}$} \quad
  \textbf{Robert Cavin$^{1}$} \quad
  \textbf{Michele A.\ Cox$^{3}$} \quad
  \textbf{Vijay Rajanna$^{3}$} \quad
  \textbf{Takumi Bolte$^{3}$}  \\
  \textbf{Melissa Hunfalvay$^{3}$} \quad
  \textbf{Ulas Bagci$^{2}$} \quad
  \textbf{Michael J.\ Proulx$^{1}$} \\[6pt]
  $^1$Meta Reality Labs Research\quad $^2$Northwestern University  \quad $^3$HarmonEyes
}

\begin{document}

\maketitle

\vspace{-20pt}
\begin{abstract}
Smart glasses with AI assistants are increasingly used in daily life. However, current systems lack awareness of the user's internal cognitive state, leaving them unable to proactively anticipate users' needs without access to cognitive load.
Existing methods for assessing cognitive load either rely on impractical sensors for lightweight eyewear or utilize eye gaze-based models that suffer from poor interpretability, and require task-specific fine-tuning, often failing to generalize across individuals.
We propose \textit{GazeMind}, a gaze-guided LLM agent framework for cognitive load assessment on smart glasses. It encodes eye-tracking data into structured representations for LLM-based reasoning and provides interpretable cognitive load predictions.
Importantly, GazeMind generalizes across scenarios without LLM fine-tuning through a novel task-guidance reasoning approach and achieves personalized adaptation by incorporating user-specific characteristics and historical references. 
To support evaluation, we introduce \textit{CogLoad-Bench}, the largest gaze-based cognitive load dataset with 152 participants, 40+ hours of multimodal data, and 10K+ real-time annotations across controlled and real-world tasks. Experiments show that GazeMind achieves state-of-the-art performance, outperforming baselines by over 20\% across all metrics.
\end{abstract}
\vspace{-10pt}

\section{Introduction}\label{intro}
\begin{figure}[t]
  \centering
  \vspace{-5pt}
  \includegraphics[width=0.7\linewidth]{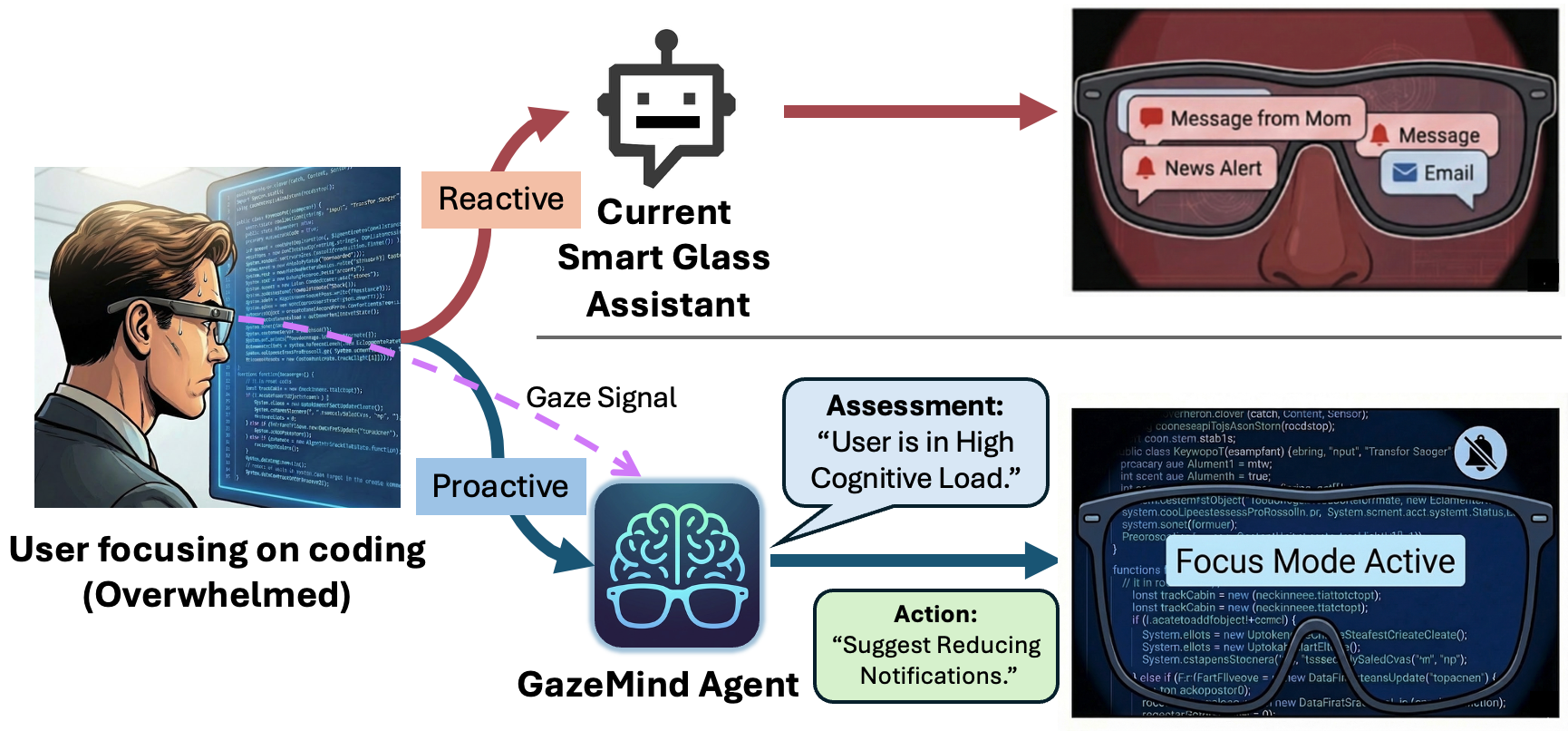}
  \caption{\textbf{Example of comparison between reactive and proactive interaction on smart glasses}. \textit{Top}: Current, reactive AI assistants are unaware of users' internal cognitive states, causing information overload and interruptions during high-demand tasks. \textit{Bottom}: Our GazeMind uses eye gaze to assess cognitive load via LLM-based reasoning, enabling proactive adaptations like suggesting Focus Mode to reduce notifications.}
  \label{teaser}
  \vspace{-25pt}
\end{figure}

Smart wearable devices, particularly smart glasses such as Ray-Ban Meta, are increasingly used in daily life  \cite{kim2021applications,rauschnabel2016augmented}. Embedded AI assistants excel at processing external context, inferring what the user is looking at \cite{wilson2025eye,wang2024gazesam}, answering user queries, and providing real-time guidance on daily tasks. However, these systems remain fundamentally blind to the user's internal cognitive state, meaning they cannot perceive the level of mental effort the user is expending while performing a task. This limitation confines current wearable AI to \textbf{reactive} interactions, responding only when explicitly prompted, rather than enabling \textbf{proactive} assistance that anticipates user needs, as shown in Figure \ref{teaser}. True proactive interaction requires awareness of the user's internal state \cite{sendhilnathan2024implicit,burlingham2024real}. Among various internal-state metrics, \textit{cognitive load}, the mental effort spent during a task \cite{plass2010cognitive}, is particularly critical. It determines whether the user can absorb additional information or requires reduced interruption. With this cognitive awareness, the AI assistant can, for instance, automatically defer notifications during periods of high mental demand or simplify its responses when the user is overwhelmed. Therefore, to build a truly human-centered, proactive wearable AI for smart glasses, enabling cognitive load assessment is imperative.

To realize cognitive load assessment on smart glasses, three fundamental questions must be addressed: what \textbf{signal} to use for sensing the user's internal state, what \textbf{model} to employ for inference, and what \textbf{data} to support the task.

\textbf{Signal.} Traditional approaches to cognitive load assessment rely on subjective self-reports or dedicated physiological sensors such as EEG and fMRI \cite{antonenko2010using,van2009tuning}. However, self-reports interrupt natural workflows, while EEG and fMRI require additional hardware, impractical for lightweight eyewear. Eye tracking offers a superior alternative. Gaze patterns, including fixations, saccades, and pupil diameter, have been found to be strong indicators for cognitive load \cite{kosch2023survey}. More crucially, eye tracking is increasingly integrated into smart glasses (e.g., Project Aria \cite{mansour2025enabling}), enabling continuous, unobtrusive data collection without additional hardware \cite{engel2023project}. Therefore, we adopt eye gaze as the primary signal for cognitive load assessment.

\textbf{Model.} Existing gaze-based cognitive load assessment methods predominantly rely on traditional machine learning methods (e.g., SVM, Random Forest) \cite{rizzo2022machine,li2024measuring,ktistakis2022colet} or deep neural networks (e.g., MLP, CNN, LSTM) \cite{wibirama2025classification,fridman2018cognitive,miles2024cogload}. 
These approaches suffer from three key challenges: (1) limited interpretability, as they predict cognitive load levels but cannot explain why certain gaze patterns indicate such levels; (2) task-specificity, which requires model fine-tuning for each new scenario; and (3) poor cross-user generalization, due to significant individual variability in gaze behavior.

Recent advances in large language models (LLMs) offer a promising solution to these challenges. LLMs encode rich prior knowledge from the cognitive science literature \cite{binz2023using}, enabling inherent understanding of relationships between gaze patterns and cognitive load, and providing semantic-level interpretability. With appropriate task guidance, LLMs can generalize across scenarios without model fine-tuning \cite{brown2020language}, interpreting identical gaze patterns differently depending on task context. Additionally, through in-context learning, LLMs support personalized adaptation by incorporating user-specific characteristics and historical references, effectively addressing individual variability without per-user fine-tuning \cite{salemi2024lamp}. Moreover, current wearable AI assistants are already built upon LLMs. Integrating cognitive load assessment into this LLM architecture enables a unified system rather than separate modules.

Building on these strengths, we propose \textbf{GazeMind}, a novel gaze-guided LLM agent framework for personalized and interpretable cognitive load assessment comprising four modules: (1) \textit{Temporal Gaze Encoding (TGE)} module converts continuous gaze signals into structured representations that LLMs can process; (2) \textit{Task-Guidance Reasoning (TGR)} module generates task-specific guidance rules, enabling LLMs to understand the gaze pattern according to different task contexts; (3) \textit{Adaptive User Profile Calibration (AUP)} module categorizes users based on their gaze behavior characteristics and dynamically adjusts interpretation according to each user's profile; (4) \textit{Cognitive Retrieval-Augmented Generation (CogRAG)} module retrieves similar historical gaze samples with their labels as in-context references, providing concrete examples to guide the LLM predictions.

\textbf{Data.} Existing gaze-based cognitive load datasets suffer from significant limitations: small user populations ($\sim$ 40 participants), controlled laboratory tasks with limited real-world applicability \cite{bhatti2025clare}, and post-session annotations where participants provide a single overall label \cite{ktistakis2022colet}, making it impossible to capture real-time cognitive load fluctuations during the task. These limitations severely impact the model's generalization and practicality. To address this, we introduce \textbf{CogLoad-Bench}, a large-scale, multi-task cognitive load dataset comprising 152 participants, 456 recordings, and over 40 hours of synchronized multimodal data, including egocentric video, audio, and eye gaze, with more than 10K real-time self-reported cognitive load annotations at three levels (low, moderate, high). The dataset includes both controlled laboratory tasks and real-world tasks, enabling evaluation of model generalization to practical scenarios.

In summary, our major contributions are as follows:

\begin{itemize}[leftmargin=1em, itemsep=0pt]
    \item \textbf{GazeMind}, a novel gaze-guided LLM agent framework for smart glasses that provides interpretable and personalized cognitive load assessment. Through task-guidance reasoning and adaptive user profile calibration, our method generalizes across diverse tasks without fine-tuning while addressing individual variability.
    
    \item \textbf{CogLoad-Bench}, largest gaze-based cognitive load dataset to date, featuring 152 participants, 40+ hours of multimodal data, and 10K+ real-time annotations across controlled and real-world tasks.

    \item Experimental results show that GazeMind achieves state-of-the-art performance, outperforming existing methods by more than \textbf{20\%} across all metrics.
\end{itemize}

\begin{figure}[t] 
  \begin{center}
    \centerline{\includegraphics[width=0.95\linewidth]{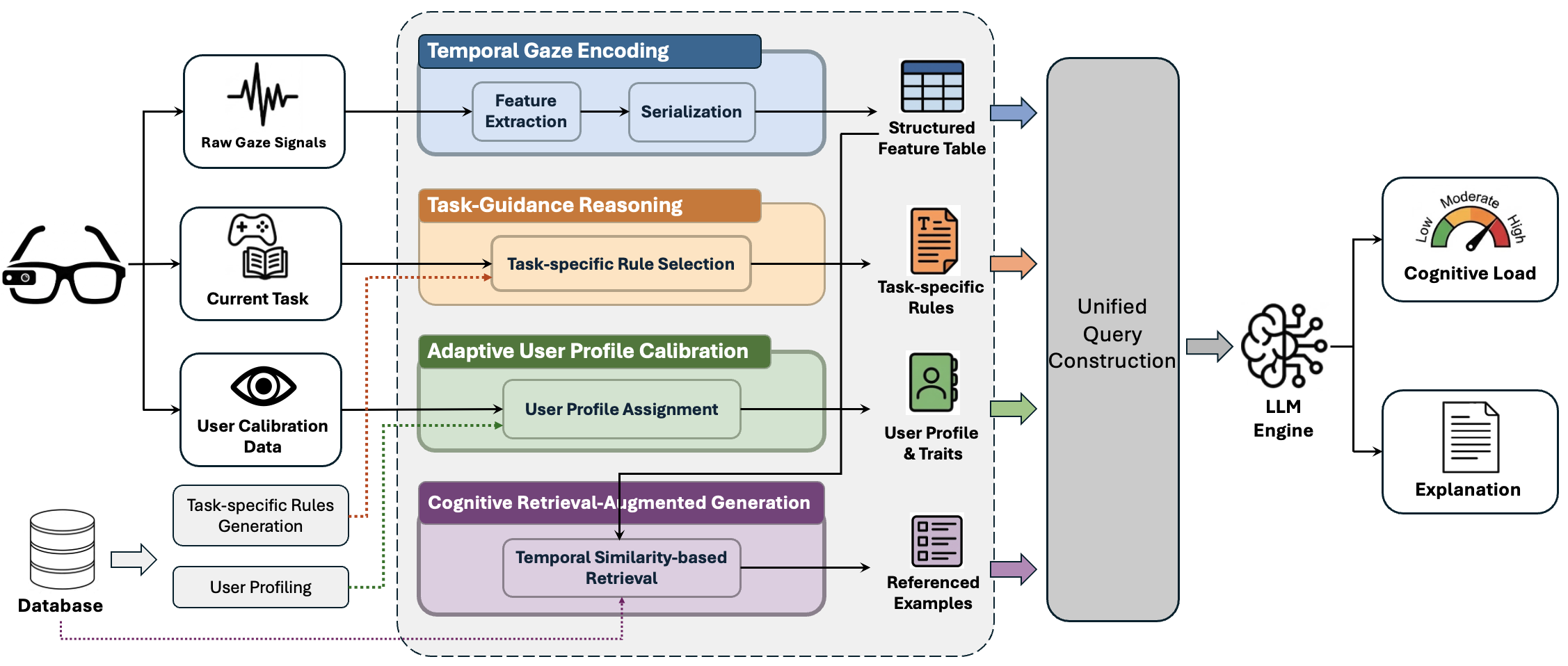}}
    \caption{Overview of our proposed \textbf{GazeMind} framework.}\label{method_fig}
  \end{center}
  \vspace{-30pt}
\end{figure}
\vspace{-6pt}
\section{Related Work}
\vspace{-6pt}
\subsection{Gaze-based Cognitive Load Assessment Methods}
Most gaze-based cognitive load assessment methods follow two stages: extracting features (e.g., fixations, saccades, blinks, pupil diameter) and classifying cognitive load via traditional machine learning methods or deep neural networks. Traditional approaches apply classifiers such as Support Vector Machines (SVM), Random Forest, and k-Nearest Neighbors on the extracted gaze features \cite{rizzo2022machine,skaramagkas2021cognitive,chen2025machine,li2024measuring,ktistakis2022colet,cury2019hybrid}, while deep learning methods adopt Convolutional Neural Networks (CNNs) \cite{miles2024cogload,nishitha2023eye}, Long Short-Term Memory networks (LSTM) \cite{rahman2021vision}, and Temporal Convolutional Networks (TCN) \cite{wibirama2025classification} to better capture temporal dynamics.  Some other work directly applies 3D-CNNs on eye region videos to classify the cognitive load \cite{fridman2018cognitive}. However, these methods lack interpretability and generalize poorly across tasks and users, as elaborated in Sec. \ref{intro}. Our framework leverages LLMs to address these issues through interpretable reasoning, task-specific guidance, and personalized adaptation.
\vspace{-6pt}
\subsection{Personalization and Task Adaptation in LLMs}
LLMs have shown remarkable capabilities in handling both task adaptation and user personalization without additional training \cite{brown2020language}. 
For task adaptation, LLMs can generalize to new scenarios through in-context learning and task-specific guidance, where task instructions guide the model to reason accordingly \cite{dong2024survey, wei2022chain}. 
For user personalization, prior work has explored two complementary strategies: constructing explicit user profiles from historical interactions to capture individual preferences \cite{salemi2024lamp}, and leveraging Retrieval-Augmented Generation (RAG) to dynamically retrieve relevant examples as personalized context \cite{lewis2020retrieval}.
These techniques have been successfully applied to text-based applications like recommendation systems and chatbots \cite{wu2024survey, achiam2023gpt}.
However, physiological signals like gaze pose a distinct challenge: the same gaze pattern can indicate different cognitive load levels depending on both the user and the task context \cite{holmqvist2011eye,steichen2014inferring}. Our framework bridges this gap with dedicated modules (TGR, AUP, and CogRAG) that jointly provide task context and user characteristics for gaze-based cognitive load interpretation.


\vspace{-6pt}
\section{Methodology}
\vspace{-6pt}
Given the eye gaze signal captured from smart glasses, our goal is to assess the user's cognitive load level. Let $G$ indicate a gaze sequence over a past time window, and let $y \in \{low, moderate, high\}$ denote the corresponding cognitive load level. Our \textbf{GazeMind} processes $G$ through four modules to predict $y$. As illustrated in Figure \ref{method_fig}, we first convert raw gaze signals into a structured feature table via Temporal Gaze Encoding (Sec. \ref{TGE}). The Task-Guidance Reasoning module then provides task-specific interpretation rules (Sec. \ref{TGR}). Adaptive User Profile Calibration adjusts predictions based on individual gaze characteristics (Sec. \ref{AUP}). Finally, Cognitive Retrieval-Augmented Generation retrieves similar samples as in-context references (Sec. \ref{CogRAG}). All components are integrated into a unified LLM query for the final prediction (Sec. \ref{unified}).

\subsection{Temporal Gaze Encoding} \label{TGE}
\paragraph{Raw Gaze Processing.} Raw gaze signals are difficult for LLMs to directly interpret and relate to cognitive load \cite{mondal2025gaze}. Therefore, we transform raw gaze signals into semantically meaningful feature representations grounded in cognitive science. Following established literature on gaze-based cognitive load indicators \cite{kosch2023survey}, we extract a set of interpretable gaze features from raw gaze signal, including fixation duration, saccade duration, saccade amplitude, fixation ratio, saccade ratio, blink count, and pupil size (See \ref{A_features} for definition and details). Hence, for each second $t$, we compute a $K$-dimensional feature vector from raw gaze data within that interval, denoted as $f_t = [f_t^{1}, f_t^{2}, ..., f_t^{K}]$, to represent user's gaze pattern at time $t$, where $K$ is the number of features we extracted. We then apply z-score normalization using population-level statistics from the database, converting raw feature values into deviations from the population-level baseline.
\vspace{-5pt}

\begin{figure}[t] 
  \begin{center}
    \centerline{\includegraphics[width=0.7\columnwidth]{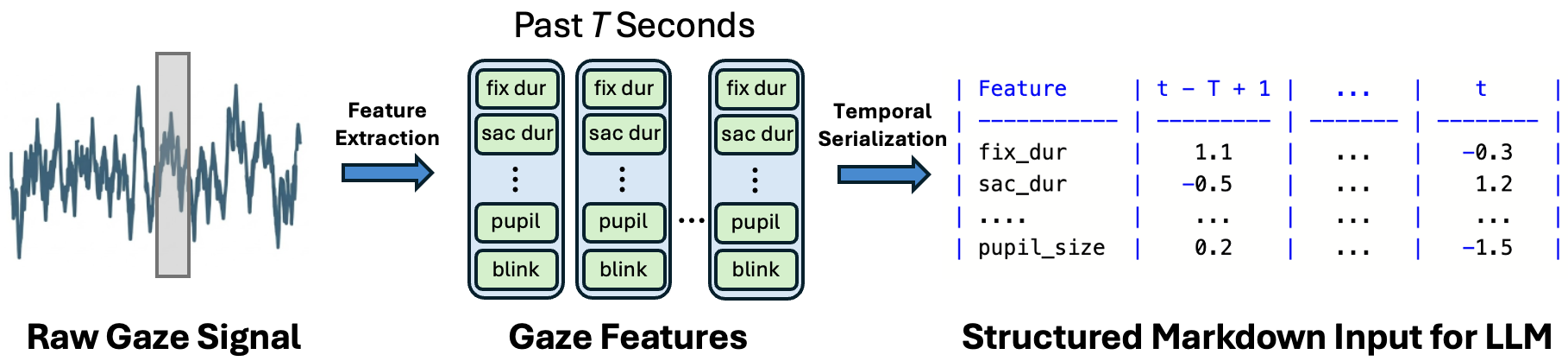}}
    \caption{\textbf{Temporal Gaze Encoding} pipeline transforming raw gaze signal into a structured markdown feature table.}\label{TGE_figure}
  \end{center}
  \vspace{-30pt}
\end{figure}

\paragraph{Temporal Feature Table.} While $f_t$ captures the gaze pattern at a single moment, cognitive load evolves dynamically during the task, and a single timestep of gaze features is insufficient to capture this temporal nature. The trend over time, whether fixation duration is progressively increasing or suddenly spiking, provides crucial context for accurate assessment. To capture these temporal dynamics, we construct a feature table that aggregates the features over the past $T$ seconds and serialize it into a structured markdown format, with rows representing features and columns indicating consecutive time steps (Figure \ref{TGE_figure}). This tabular representation preserves temporal ordering for pattern identification (e.g., rising pupil size, declining blink rate) while providing a structured format that LLMs can naturally parse, enabling the model to interpret both magnitude and trend of gaze features.

\vspace{-5pt}
\subsection{Task-Guidance Reasoning} \label{TGR}
\vspace{-5pt}
Identical gaze patterns may indicate different cognitive load levels in different task contexts. For instance, shorter fixation durations reflect low cognitive load in reading (skimming easy text), but signal high cognitive load in gaming (rapidly tracking multiple dynamic targets) \cite{kosch2023survey}. Simply providing feature definitions to the LLM leads to poor performance, as the model lacks task-specific knowledge to interpret gaze patterns accordingly. To address this, we propose the TGR module, which automatically generates task-specific interpretation rules offline and selects them during inference; rules are generated in two stages:
\vspace{-6pt}

\paragraph{Feature Distribution Analysis.} For each task, labeled gaze samples are grouped by cognitive load level (low/moderate/high), and statistics of each gaze feature are computed for each group, including mean, standard deviation, median, and quartiles ($Q_{25}, Q_{75}$). These statistics reveal whether a feature exhibits distinct distributions across cognitive load levels. If the distributions are well-separated, the feature is discriminative for that task, but if they overlap substantially, the feature is less informative.
\vspace{-6pt}

\paragraph{Rules Generation via LLM.} These statistics are then provided to an LLM, which analyzes the relationship between gaze features and cognitive load levels.  The LLM generates a structured guidance rule that specifies: (1) the most discriminative features for the task, ranked by importance; (2) the directional relationship between each feature and cognitive load; and (3) quantitative thresholds derived from the statistics. During inference, given the current task context, the corresponding rule is automatically selected and incorporated into the LLM query, guiding the interpretation of the observed gaze patterns. Note, the rules generated are consistent with established cognitive science literature, see Appendix~\ref{A_Interpretability} for a detailed validation.
\vspace{-5pt}
\subsection{Adaptive User Profile Calibration} \label{AUP}
Even with task-specific guidance, individual variability in gaze behavior remains a challenge. Users exhibit distinct baseline gaze patterns. Some naturally have a larger pupil size than others, regardless of cognitive load. Applying uniform interpretation rules to all users leads to errors. We address this through AUP, which categorizes users into profiles and adjusts interpretation accordingly. Our approach requires only a brief calibration phase, avoiding the need for per-user model fine-tuning.
\vspace{-2pt}

\paragraph{User Profiling.} We characterize each user's gaze behavior by four aspects: blink intensity (mean blink count), pupil sensitivity (standard deviation of pupil size), pupil size baseline (mean pupil size), and gaze instability (spatial dispersion of gaze points). We aggregate these features across all samples for each user in the database and apply clustering \cite{hartigan1979algorithm}, which revealed three well-separated user profiles:
\vspace{-5pt}

\begin{itemize}[leftmargin=1em, itemsep=0pt]
    \item \textbf{High-Reactor}: large pupil size baseline with high pupil sensitivity; normal blink rate and gaze instability.
    \item \textbf{Low-Reactor}: low pupil sensitivity, blink rate, and gaze instability.
    \item \textbf{Restless}: twice the blink intensity of others with the highest gaze instability.
\end{itemize}
\vspace{-5pt}

During inference, for a new user, a short calibration session that includes different cognitive load levels is first collected. The user's feature characteristics are computed and matched to the closest profile. We also calculate the user's own personalized baselines (pupil size, blink count), which are extracted directly from their calibration data. The assigned profile description, traits, and the user's own gaze baselines are then incorporated into the LLM query, instructing the model to adjust its interpretation accordingly. For instance, if a user is classified as Low-Reactor, the LLM is informed that this user shows naturally reduced gaze variability, and therefore smaller deviations from baseline may still indicate significant cognitive load changes. See \ref{A_user} for more details.
\vspace{-5pt}

\subsection{Cognitive Retrieval-Augmented Generation} \label{CogRAG}
Task-specific rules and user profiles provide general guidance, but without concrete examples, the LLM lacks direct reference points for comparison. Hence, we propose CogRAG, which retrieves similar gaze samples with labels from a database to serve as in-context references, providing LLM with concrete examples to guide its prediction. Unlike traditional RAG \cite{lewis2020retrieval} that relies on learned embeddings to create dense vectors for retrieval, which is computationally expensive, CogRAG avoids this by directly using temporal statistical descriptors as lightweight embeddings for similarity matching, enabling efficient real-time inference on smart glasses.

We first construct a retrieval database containing structured temporal feature tables paired with cognitive load labels. Each sample is tagged with its task category and user profile. During retrieval, only samples from the same task category and user profile are considered, since gaze patterns carry different meanings across tasks and user types.
\vspace{-5pt}

\paragraph{Temporal Similarity Metric.} Standard RAG approaches typically compute similarity using metrics like cosine or Euclidean distance on raw vectors \cite{borgeaud2022improving}. However, such metrics are unsuitable for our temporal feature tables $\mathbf{X} \in \mathbb{R}^{T\times K}$, as element-wise comparison treats each timestep independently and fails to capture temporal patterns such as trending upward or remaining stable. To address this, we map each feature table into a statistical vector. For each feature $k\in \{1,...,K\}$, we compute the mean $\mu_k$ to represent the overall magnitude, the standard deviation $\sigma_k$ to capture variability, and trend $\beta_k$, linear slope over time, to reflect temporal direction. By concatenating these descriptors, we construct a statistical vector $\Phi(\mathbf{X})=[\mu_1, \sigma_1, \beta_1, \mu_2, \sigma_2, \beta_2, ..., \mu_K, \sigma_K, \beta_K]$. The similarity between two samples $\mathbf{X}_i$ and $\mathbf{X}_j$ is then defined by
\begin{equation}
\mathcal{D}(\mathbf{X}_i, \mathbf{X}_j) = \|\Phi(\mathbf{X}_i) - \Phi(\mathbf{X}_j)\|_2.
\end{equation}
This calculation is also more robust to noise compared to element-wise comparison.

Once retrieved, the feature tables along with their cognitive load labels are formatted as reference examples in the LLM query, indicating how similar gaze patterns correspond to certain cognitive load levels. Note that during inference, the LLM is instructed to qualitatively validate retrieved examples with the current sample to check if gaze pattern is consistent before using them.

\subsection{Unified Inference} \label{unified}
\vspace{-5pt}
During inference, all components are integrated into a single LLM query. Given a test sample, the query includes: (1) the temporal feature table from TGE; (2) the task-specific guidance rule from TGR; (3) the user profile description and traits from AUP; and (4) the retrieved reference examples from CogRAG. The LLM processes all information jointly and outputs both the predicted cognitive load level and a natural language explanation justifying the prediction based on the observed gaze patterns. For prompt details, see \ref{A_prompt}.

\vspace{-6pt}
\section{CogLoad-Bench Dataset}
\vspace{-6pt}

\begin{wrapfigure}{r}{0.5\linewidth}
  \centering
  \vspace{-10pt}
  \includegraphics[width=0.93\linewidth]{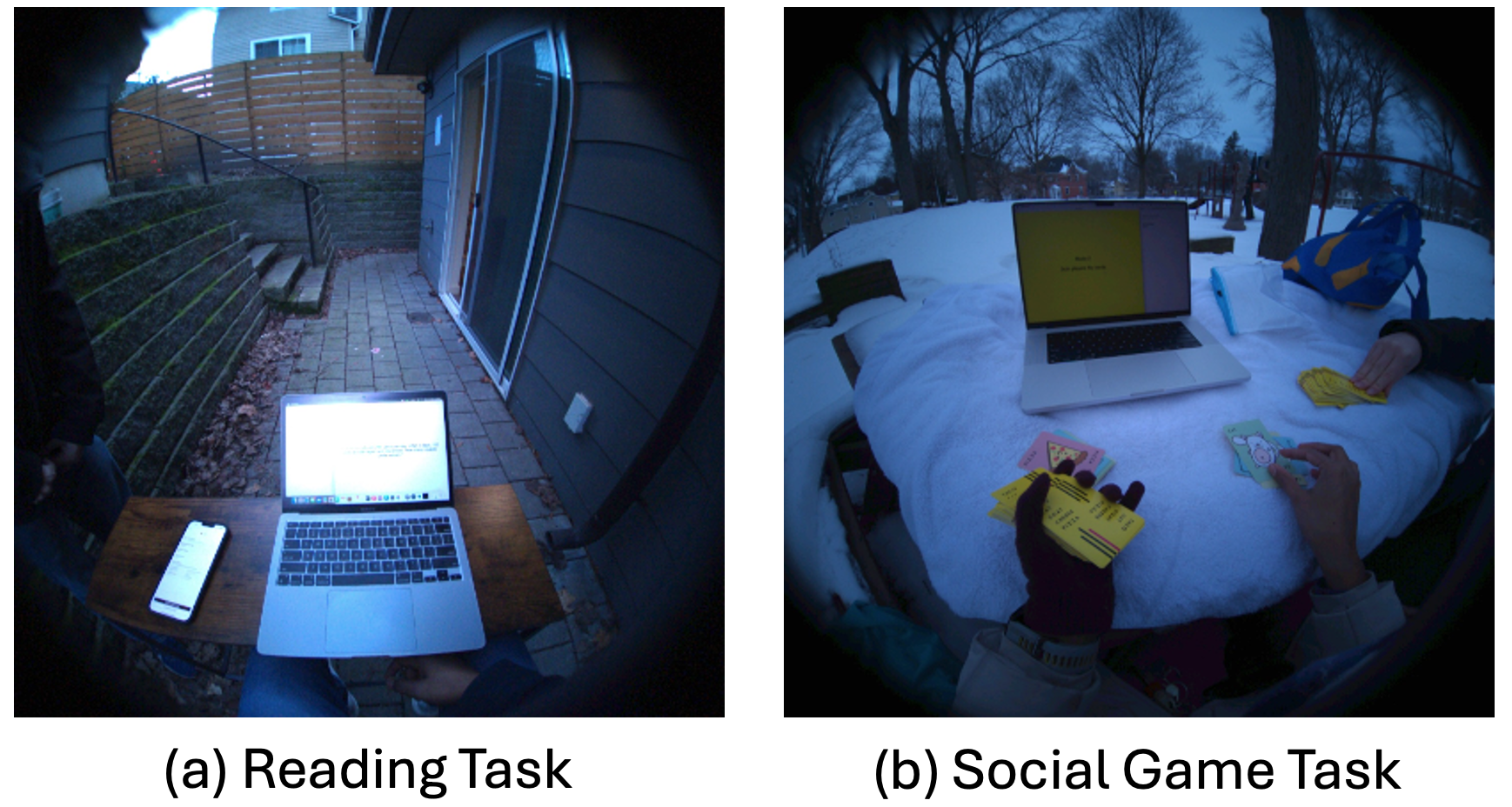}
  \caption{Example egocentric views from \textbf{CogLoad-Bench}. (a) Reading Task. (b) Social Game Task.}
  \label{task_figure}
  \vspace{-20pt}
\end{wrapfigure}

Existing gaze-based cognitive load datasets \cite{bhatti2025clare, pillai2020response, chiossi2024understanding, angkan2024multimodal} have a number of disadvantages, including limited user population, lack of real-time annotations, and over-reliance on controlled laboratory tasks, making them insufficient for developing generalized models for smart glasses. In contrast, CogLoad-Bench (Figure \ref{task_figure}) offers several distinctive advantages:

\vspace{-5pt}
\begin{itemize}[leftmargin=1em, itemsep=0pt]
    \item \textbf{Large User Population.} With 152 participants involved in the dataset, CogLoad-Bench is approximately 4x larger than the existing datasets from the aspect of the user group. This enables robust evaluation of cross-user generalization required for wearable assistant model.
    \item \textbf{Real-time Annotations.} Unlike post-session labeling used in existing datasets that provides a single cognitive load rating for the whole recording, we collect in-task labels every 15–30 seconds, generating over 10K temporally-aligned annotations that capture cognitive load fluctuations throughout task execution.
    \item \textbf{Diverse Tasks.} Beyond the controlled laboratory paradigms of prior work, CogLoad-Bench includes both controlled and real-world tasks, enabling rigorous benchmarking and evaluation of practical generalization.
\end{itemize}

\vspace{-5pt}
\subsection{Tasks}
\vspace{-5pt}
We designed three tasks spanning controlled and real-world scenarios for the participants: a reading-based reasoning task, an interactive social game with environmental distractions, and an audio listening task with clear discriminative difficulty level transitions as user calibration data for getting user gaze behavior under different mental efforts. To further enhance real-world validity, data collection is in naturalistic environments rather than controlled laboratory spaces. This combination of diverse tasks and realistic settings enables both rigorous benchmarking and evaluation of real-world generalization.
\vspace{-5pt}
\paragraph{Reading Task.} Participants read and solved reasoning problems, responding verbally without external aids. Problems included temporal reasoning (``What day is two days after Monday?") and mental arithmetic (``If you have 20 and spend 7..."), each at three difficulty levels based on complexity. Each level lasted approximately one minute, with difficulty cycling through a fixed progression to induce cognitive load variations.
\vspace{-5pt}

\paragraph{Social Game Task.} Participants played an interactive social game based on a modified version of \emph{Taco Cat Goat Cheese Pizza} with a partner. Task difficulty was manipulated across three levels by varying gameplay structure and distraction: in the easy condition, the game followed a fixed sequence with no competing player; in the medium condition, participants alternated turns with the partner; and in the hard condition, unpredictable verbal cues or background auditory distraction were introduced. Each level lasted about one minute, cycling through fixed progressions.
\vspace{-5pt}

\paragraph{Audio Task.} Participants listened to a sequence of spoken letters and verbally responded when the current letter matched one from $N$ positions earlier ($N\in\{1,2,3\}$). Higher $N$ values require tracking more letters in memory, inducing higher cognitive load. Blocks cycled through difficulty levels in fixed progressions. This standardized paradigm serves as calibration data for user profile construction, allowing the system to learn individual gaze characteristics (e.g., pupil sensitivity, blink patterns) under controlled cognitive load variations before generalizing to real-world tasks.

\begin{table}[t]
  \caption{Performance comparison of GazeMind against baseline models. We categorize methods by their learning paradigms: Supervised, Zero-shot, and In-Context Learning (ICL).}
  \label{tab:main_results}
  \centering
  \resizebox{0.8\linewidth}{!}{%
  \begin{tabular}{lccccc}
    \toprule
    \textbf{Methods} & \textbf{Paradigm} & \textbf{Accuracy $\uparrow$} & \textbf{Precision $\uparrow$} & \textbf{Recall $\uparrow$} & \textbf{F1 $\uparrow$} \\
    \midrule
    Decision Tree    & Supervised & 33.17 & 33.23 & 33.10 & 32.80 \\
    SVM              & Supervised & 36.62 & 34.89 & 34.86 & 34.12 \\
    MLP              & Supervised & 37.39 & 35.13 & 34.92 & 34.34 \\
    TCN              & Supervised & 36.31 & 35.68 & 35.78 & 35.58 \\
    Transformer      & Supervised & 37.16 & 34.55 & 34.40 & 33.15 \\
    LSTM             & Supervised & 37.96 & 35.99 & 35.87 & 35.48 \\
    Time-LLM         & Supervised & 38.11 & 34.78 & 34.72 & 33.60 \\
    \midrule
    Llama-3.3-70B    & Zero-shot  & 39.30 & 37.11 & 37.05 & 36.82 \\
    GPT-4o           & Zero-shot  & \underline{39.62} & \underline{40.81} & \underline{40.61} & \underline{38.98} \\
    \midrule
    \textbf{GazeMind} & \textbf{ICL} & \textbf{62.73} & \textbf{62.58} & \textbf{61.77} & \textbf{62.11} \\
    \bottomrule
  \end{tabular}}
  \vspace{-20pt}
\end{table}

\vspace{-7pt}
\subsection{Data Acquisition}
All data were recorded using Project Aria glasses \cite{engel2023project}, capturing synchronized eye gaze at 90 Hz, egocentric RGB video at 10 Hz, and audio at 48 kHz. Each participant contributed approximately 15 minutes of task data, with 5 minutes for each task. In this way, we collect 456 recordings with over 40 hours of multimodal data across 152 participants in total.

\vspace{-5pt}
\subsection{Annotation Protocol}
\vspace{-3pt}
Participants verbally reported their perceived cognitive load every 15–30 seconds during task execution using a 7-level ordinal scale (very low, low, moderate-low, moderate, moderate-high, high, very high), which was collapsed into three levels for modeling: Low (very low, low, moderate-low), Moderate, and High (moderate-high, high, very high).  This aggregation reduces boundary ambiguity and class imbalance while preserving meaningful distinctions. Labels are extracted by transcribing the verbal reports using the speech-to-text tool \cite{radford2023robust} and manually verified on representative samples. To enable time-series modeling, annotations were back-propagated between consecutive report points to produce per-second labels. See Appendix~\ref{A_data} for more dataset details.

\vspace{-8pt}
\section{Experiment}
\vspace{-5pt}

We evaluate GazeMind on the proposed CogLoad-Bench against baselines commonly used in gaze-based cognitive load assessment~\cite{ktistakis2022colet,rizzo2022machine,rahman2021vision,wibirama2025classification}: Decision Tree, SVM, MLP, TCN, Transformer, LSTM, and Time-LLM~\cite{jin2023time}, a strong time-series baseline. All methods use the same past 5-second temporal features. We also evaluate Llama-3.3-70B \cite{grattafiori2024llama} and GPT-4o with TGE-encoded gaze features and definitions, but without task-guidance reasoning, user profiles, or retrieved examples. Comparison implementation details are in Appendix \ref{A_implementation}.

\vspace{-5pt}
\subsection{Experiment Setup}
\vspace{-5pt}

\paragraph{Dataset Splits.} We partition CogLoad-Bench using a cross-user split to evaluate generalization to unseen users. Specifically, we randomly assign 70\% of participants (106 users) for training or database construction, 30\% (46 users) for testing. This ensures no overlap between users in training and evaluation, reflecting realistic deployment scenarios where system must generalize to new users.

\textbf{Evaluation Metrics.} We report four standard classification metrics, including Accuracy, Precision, Recall, and F1-score, with the latter three computed using macro-averaging.

\paragraph{Implementation Details.} GazeMind uses GPT-4o \cite{achiam2023gpt} as the backbone LLM with temperature set to 0 for reproducible predictions. The temporal window $T$ is set to 5 seconds, capturing sufficient temporal dynamics while maintaining real-time responsiveness suitable for smart glasses deployment. 
Proposed modules only use the training set as the database (no testing user data).
Task-specific rules are automatically pre-generated for each task category. The audio task is considered user calibration data because it provides controlled cognitive load variations for characterizing individual gaze behavior. For CogRAG, we retrieve 3 reference examples from the database within the same task and user profile group as the current sample.

\subsection{Main Results}
\begin{figure}[t]
  \centering
  \begin{minipage}{0.48\linewidth}
    \centering
    \centerline{\includegraphics[width=\linewidth]{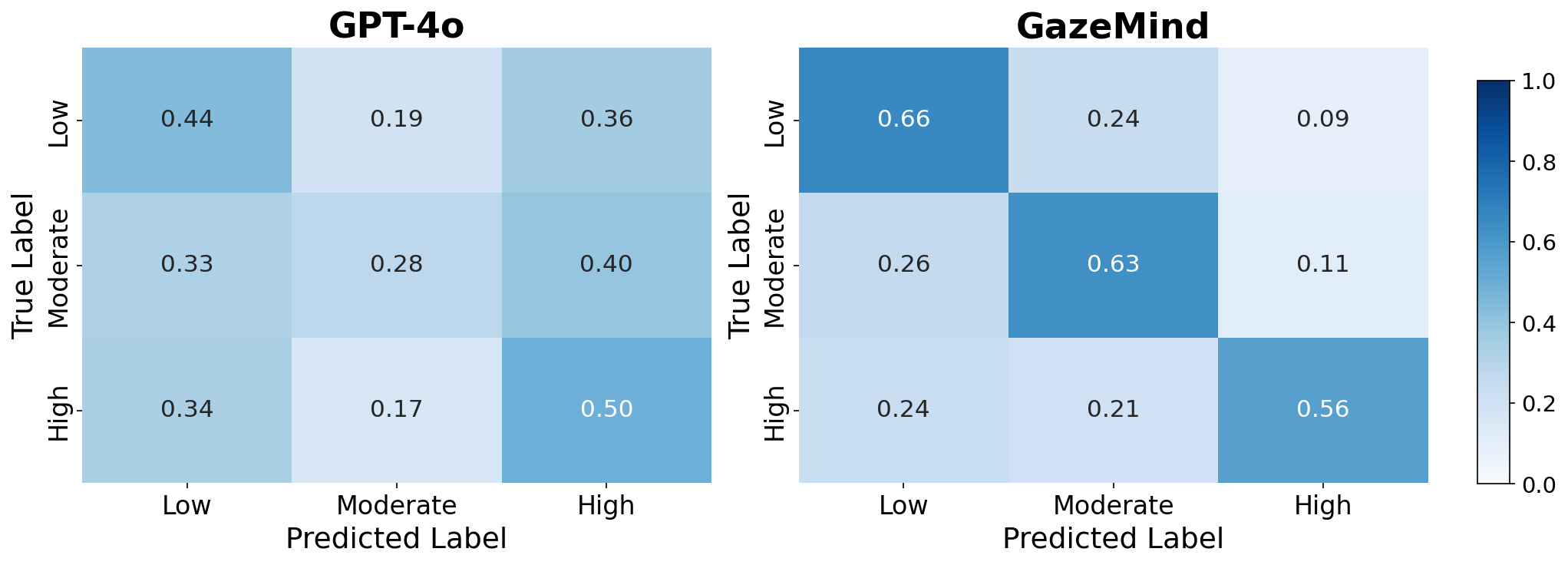}}
    \caption{Confusion Matrix Comparison between GPT-4o and GazeMind.}\label{confisuion_fig}
  \end{minipage}
  \hfill
  \begin{minipage}{0.48\linewidth}
    \centering
    \centerline{\includegraphics[width=\linewidth]{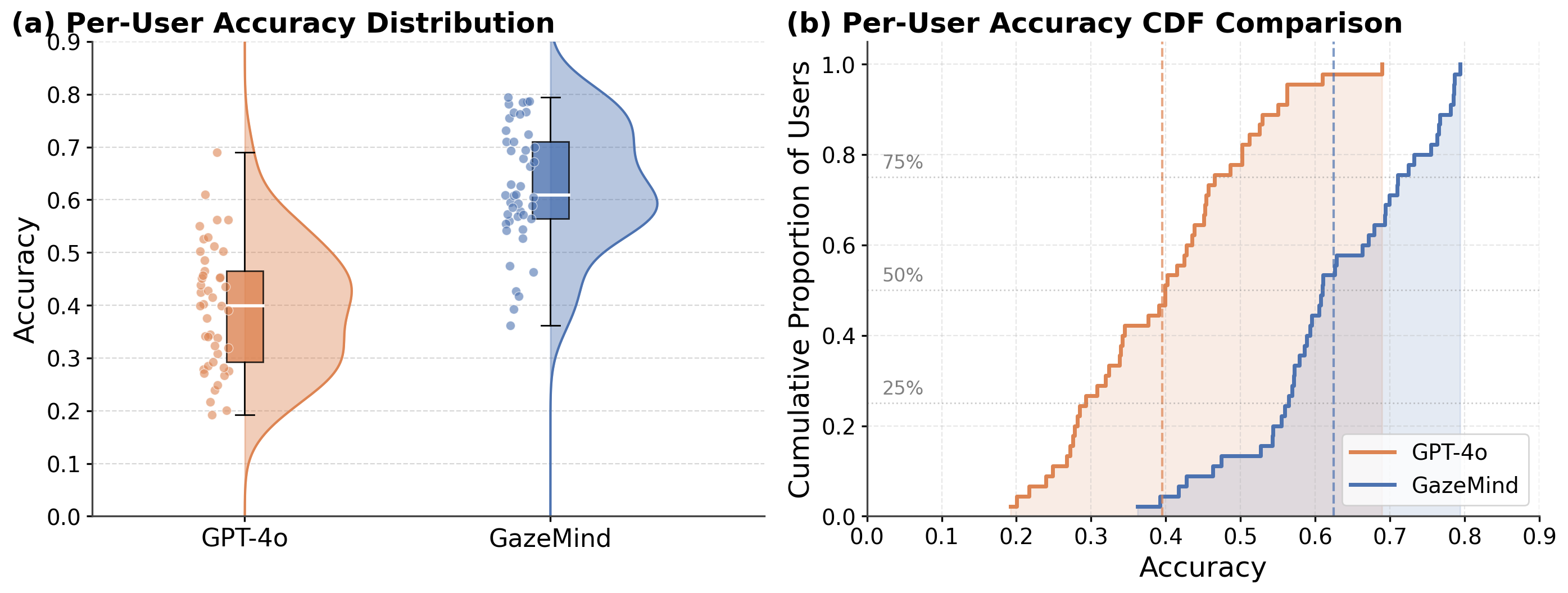}}
    \caption{Analysis of (a) per-user accuracy distributions, and (b) cumulative distribution comparison.}\label{peruser_fig}
  \end{minipage}
  \vspace{-20pt}
\end{figure}

Table \ref{tab:main_results} shows the overall model performance comparison. GazeMind achieves 62.73\% accuracy and 62.11\% F1-score, outperforming all baselines by over 20\%. Supervised methods achieve only 33-38\% accuracy, indicating poor cross-user and cross-task generalization. Zero-shot LLMs show only marginal improvement but lack sufficient task-specific and user-specific knowledge. The confusion matrices in Figure \ref{confisuion_fig} further illustrate this: GPT-4o shows bias toward predicting High, while GazeMind achieves balanced predictions across all three classes.

\begin{wraptable}{r}{0.42\linewidth}
\vspace{-12pt}
\centering
\small
\caption{Task performance for Reading and Gaming scenarios.}
\label{tab:task_results}
\resizebox{\linewidth}{!}{%
\begin{tabular}{lcccc}
\toprule
& \multicolumn{2}{c}{\textbf{Reading}} & \multicolumn{2}{c}{\textbf{Gaming}} \\
\cmidrule(r){2-3} \cmidrule(l){4-5}
\textbf{Methods} & F1 $\uparrow$ & Acc $\uparrow$ & F1 $\uparrow$ & Acc $\uparrow$ \\
\midrule
Decision Tree    & 33.55 & 34.51 & 30.92 & 31.93 \\
SVM              & 31.83 & 38.76 & 31.87 & 34.63 \\
MLP              & 32.42 & 37.53 & 33.07 & 37.26 \\
LSTM             & 31.68 & 38.60 & 31.19 & 37.37 \\
Llama-3.3-70B    & 37.97 & 40.66 & 35.66 & 38.02 \\
GPT-4o           & 39.85 & 43.32 & 36.06 & 36.15 \\
\midrule
\textbf{GazeMind} & \textbf{63.96} & \textbf{64.98} & \textbf{59.64} & \textbf{60.63} \\
\bottomrule
\end{tabular}}
\vspace{-15pt}
\end{wraptable}

\paragraph{Task Generalization.} Table \ref{tab:task_results} shows GazeMind outperforms baselines on both Reading and Gaming tasks. This strong cross-task performance is attributed to the TGR module, which explicitly informs LLM how gaze patterns relate to cognitive load in each task context, enabling more accurate interpretation. Note that reading yields higher performance, likely due to more stable gaze patterns compared to the Social Game task with more environmental distractions. More results see \ref{A_task}.

\vspace{-5pt}
\paragraph{Personalization Adaptation.} Figure \ref{peruser_fig} analyzes per-user accuracy distribution to evaluate personalization effectiveness. As shown in Figure \ref{peruser_fig}(a), most users achieve above 60\% accuracy with GazeMind, whereas nearly half of users fall below 40\% with GPT-4o. The cumulative distribution in Figure \ref{peruser_fig}(b) further shows that GazeMind shifts the entire distribution rightward rather than benefiting only a subset of users. Notably, the lower tail is substantially lifted, indicating that even low-performing users benefit from our approach. This demonstrates that the proposed AUP module effectively addresses individual variability.

More analyses are in appendix: source-of-gain comparisons (\ref{A_source}), latency and on-device deployment (\ref{A_time}), robustness to label noise, missing data (\ref{A_Robustness}), model interpretability (\ref{A_Interpretability}), label reliability (\ref{A_label}), cross-LLM generalization (\ref{A_cross}), cross-task generalization (\ref{A_crosstask}), and temporal window ablation (\ref{A_ablationtemp}).

\vspace{-5pt}
\subsection{Ablation Study}

\begin{wraptable}{r}{0.55\linewidth}
\vspace{-12pt}
\centering
\small
\caption{Ablation study on CogLoad-Bench. We show the cumulative contribution of each module: TGE, TGR, AUP, and CogRAG.}
\label{tab:ablation}
\resizebox{\linewidth}{!}{%
\begin{tabular}{lcccc|cc}
\toprule
 & TGE & TGR & AUP & CogRAG & \textbf{Acc $\uparrow$} & \textbf{F1 $\uparrow$} \\
\midrule
GPT-4o    & \checkmark & --         & --         & --         & 39.62 & 38.98 \\
+ TGR     & \checkmark & \checkmark & --         & --         & 45.34 & 43.70 \\
+ AUP     & \checkmark & \checkmark & \checkmark & --         & 49.10 & 47.14 \\
+ CogRAG  & \checkmark & \checkmark & \checkmark & \checkmark & \textbf{62.73} & \textbf{62.11} \\
\bottomrule
\end{tabular}}
\vspace{-10pt}
\end{wraptable}

Table \ref{tab:ablation} presents the contribution of each module. Adding TGR improves accuracy by providing task-specific guidance on how gaze patterns relate to cognitive load in different contexts. AUP yields an additional gain, showing the importance of personalized adaptation that accounts for individual variability in gaze behavior. CogRAG provides largest improvement, as it retrieves reference examples from the same task category and user profile, providing highly representative comparison cases that share similar gaze-cognitive load relationships with the current sample.
\vspace{-5pt}

\subsection{Case Study}
\begin{figure}[t] 
  \begin{center}
    \centerline{\includegraphics[width=\linewidth]{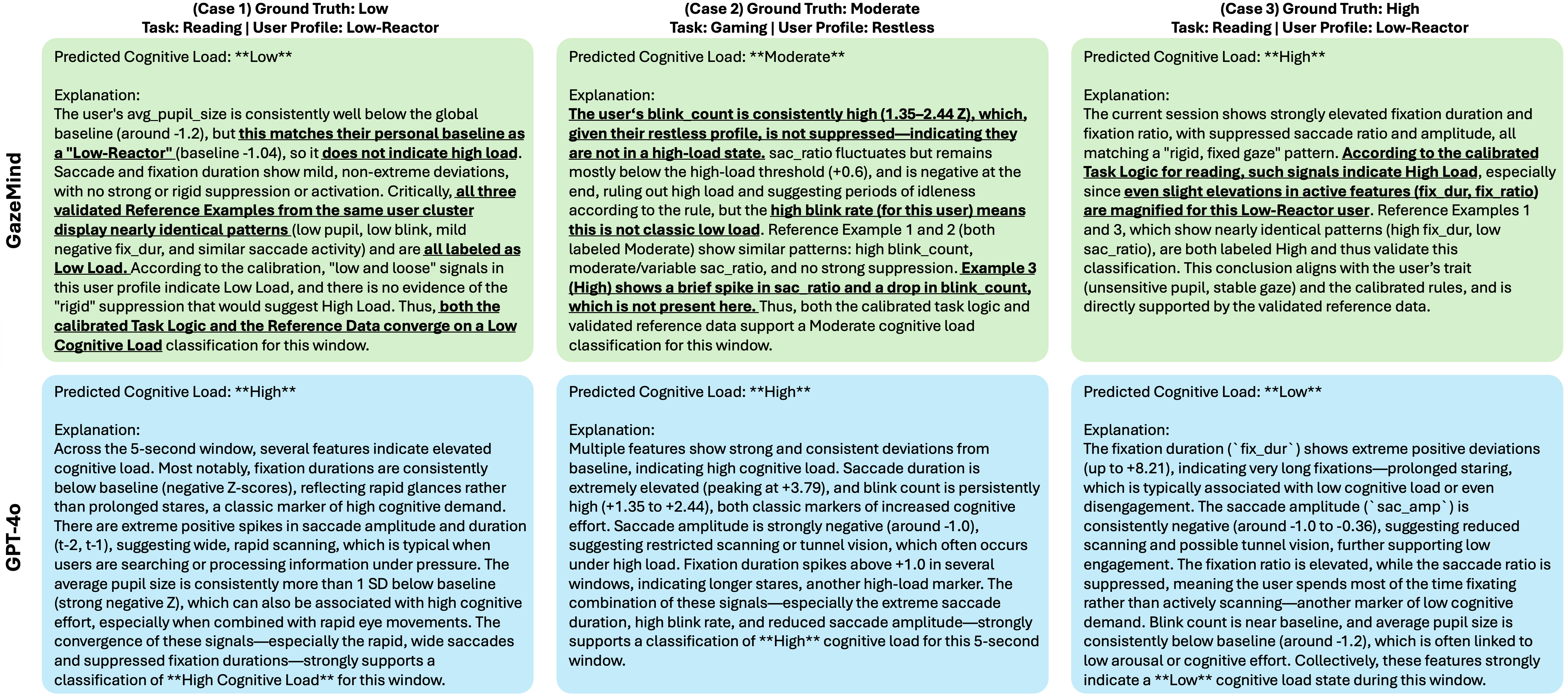}}
    \caption{\textbf{Case study comparing GazeMind and GPT-4o.} GazeMind (top) correctly predicts all three cases by incorporating user-specific characteristics (Low-Reactor, Restless user profiles), task-guidance reasoning, and reference examples, while GPT-4o (bottom) misclassifies all cases because it fails to capture individual and task-specific variations in gaze behavior.}\label{case_figure}
  \end{center}
  \vspace{-30pt}
\end{figure}

Figure \ref{case_figure} compares GazeMind and GPT-4o on three representative cases, illustrating how our modules enable accurate predictions. In Case 1, the user's pupil size is consistently below the population baseline. GPT-4o misinterprets this as ``high cognitive effort", while GazeMind correctly recognizes that this Low-Reactor user has a naturally low pupil baseline, so the deviation does not indicate high load. Both the calibrated task logic and retrieved reference examples converge on a Low classification. In Case 2, GazeMind observes the user's Restless profile with naturally high blink rate. Notably, while CogRAG retrieves Example 3 labeled as High, GazeMind identifies that its gaze pattern differs from the current sample and excludes it from consideration, showing its ability to recheck the retrieved samples instead of purely trusting them. In contrast, GPT-4o analyzes features independently without integrating user context, leading to misclassification. 
In Case 3, GazeMind strictly follows the task guidance that slight elevations in fixation duration and fixation ratio indicate High load for reading tasks, especially for Low-Reactor users. The reasoning aligns precisely with our design: combining task-specific rules with user profile calibration, and validated by reference examples.

\vspace{-10pt}
\section{Discussion} \label{discussion}
\vspace{-6pt}

\paragraph{Insights.} Our results show that LLMs can effectively extract patterns from time-series data like gaze features, but achieving robust classification performance requires sufficient context. This is evidenced by each proposed modules' performance gain, where increased contextual information directly improved LLM reasoning.
Importantly, our framework offers a generalizable template for applying LLMs to other time-series domains such as physiological signal processing on wearables, where similar challenges of task-dependent interpretation and individual variability commonly exist.
We have also discussed further impact statement, ethical and privacy issues at Appendix \ref{A_privacy}.
\vspace{-5pt}

\paragraph{Limitations.} First, although GazeMind avoids model fine-tuning for each task category, task-guidance rules still need to be derived. Second, self-reported cognitive load labels carry inherent subjectivity, limits current accuracy to human agreement ceiling (Appendix~\ref{A_label}).
Third, our approach relies solely on gaze data, but additional modalities would improve contextual understanding.
\vspace{-5pt}

\paragraph{Future Work.} 
First, we aim to extend rule generation to an online, zero-shot setting, leveraging planning-based reasoning to infer rules directly from task descriptions without offline analysis.
Second, we will include other modalities such as egocentric video and audio to enrich the understanding of user behavior and real-world context, which also help reduce the noise in self-reported cognitive load labels.
Finally, to realize the proactive interaction in Figure \ref{teaser}, an action-oriented LLM agent should add after GazeMind to trigger real-time functionality on smart glasses.
For example, the agent can proactively delay non-critical notifications if user is experiencing high cognitive load.

\vspace{-10pt}
\section{Conclusion}
\vspace{-6pt}

We present GazeMind, a gaze-guided LLM agent framework for personalized cognitive load assessment on smart glasses. By integrating TGE, TGR, AUP, and CogRAG modules, GazeMind enables LLMs to interpret gaze patterns with task-specific and user-specific context, achieving interpretable predictions without the need for model fine-tuning. We also introduce CogLoad-Bench, the largest gaze-based cognitive load dataset with 152 participants and 10K+ real-time annotations. Experiments demonstrate that GazeMind outperforms baselines by over 20\% across all metrics. This work provides a foundation for building proactive, human-centered AI assistants on smart glasses.

\newpage
\bibliographystyle{plainnat}
\bibliography{references}

@article{kim2021applications,
  title={Applications of smart glasses in applied sciences: A systematic review},
  author={Kim, Dawon and Choi, Yosoon},
  journal={Applied Sciences},
  volume={11},
  number={11},
  pages={4956},
  year={2021},
  publisher={MDPI}
}

@article{rauschnabel2016augmented,
  title={Augmented reality smart glasses: An investigation of technology acceptance drivers},
  author={Rauschnabel, Philipp A and Ro, Young K},
  journal={International Journal of Technology Marketing},
  volume={11},
  number={2},
  pages={123--148},
  year={2016},
  publisher={Inderscience Publishers (IEL)}
}

@article{plass2010cognitive,
  title={Cognitive load theory},
  author={Plass, Jan L and Moreno, Roxana and Br{\"u}nken, Roland},
  year={2010},
  publisher={Cambridge university press}
}

@article{kosch2023survey,
  title={A survey on measuring cognitive workload in human-computer interaction},
  author={Kosch, Thomas and Karolus, Jakob and Zagermann, Johannes and Reiterer, Harald and Schmidt, Albrecht and Wo{\'z}niak, Pawe{\l} W},
  journal={ACM Computing Surveys},
  volume={55},
  number={13s},
  pages={1--39},
  year={2023},
  publisher={ACM New York, NY}
}

@article{antonenko2010using,
  title={Using electroencephalography to measure cognitive load},
  author={Antonenko, Pavlo and Paas, Fred and Grabner, Roland and Van Gog, Tamara},
  journal={Educational psychology review},
  volume={22},
  number={4},
  pages={425--438},
  year={2010},
  publisher={Springer}
}

@article{engel2023project,
  title={Project aria: A new tool for egocentric multi-modal ai research},
  author={Engel, Jakob and Somasundaram, Kiran and Goesele, Michael and Sun, Albert and Gamino, Alexander and Turner, Andrew and Talattof, Arjang and Yuan, Arnie and Souti, Bilal and Meredith, Brighid and others},
  journal={arXiv preprint arXiv:2308.13561},
  year={2023}
}

@article{van2009tuning,
  title={Tuning down the emotional brain: an fMRI study of the effects of cognitive load on the processing of affective images},
  author={Van Dillen, Lotte F and Heslenfeld, Dirk J and Koole, Sander L},
  journal={Neuroimage},
  volume={45},
  number={4},
  pages={1212--1219},
  year={2009},
  publisher={Elsevier}
}

@inproceedings{fridman2018cognitive,
  title={Cognitive load estimation in the wild},
  author={Fridman, Lex and Reimer, Bryan and Mehler, Bruce and Freeman, William T},
  booktitle={Proceedings of the 2018 chi conference on human factors in computing systems},
  pages={1--9},
  year={2018}
}

@article{rizzo2022machine,
  title={A machine learning approach for detecting cognitive interference based on eye-tracking data},
  author={Rizzo, Antonio and Ermini, Sara and Zanca, Dario and Bernabini, Dario and Rossi, Alessandro},
  journal={Frontiers in Human Neuroscience},
  volume={16},
  pages={806330},
  year={2022},
  publisher={Frontiers Media SA}
}

@article{li2024measuring,
  title={Measuring and classifying students' cognitive load in pen-based mobile learning using handwriting, touch gestural and eye-tracking data},
  author={Li, Qingchuan and Luximon, Yan and Zhang, Jiaxin and Song, Yao},
  journal={British Journal of Educational Technology},
  volume={55},
  number={2},
  pages={625--653},
  year={2024},
  publisher={Wiley Online Library}
}

@article{ktistakis2022colet,
  title={COLET: A dataset for COgnitive workLoad estimation based on eye-tracking},
  author={Ktistakis, Emmanouil and Skaramagkas, Vasileios and Manousos, Dimitris and Tachos, Nikolaos S and Tripoliti, Evanthia and Fotiadis, Dimitrios I and Tsiknakis, Manolis},
  journal={Computer Methods and Programs in Biomedicine},
  volume={224},
  pages={106989},
  year={2022},
  publisher={Elsevier}
}

@article{wibirama2025classification,
  title={Classification of cognitive load using deep learning based on eye movement indices},
  author={Wibirama, Sunu and Alfarozi, Syukron Abu Ishaq and Suhari, Ahmad Riznandi and Fristiana, Ayuningtyas Hari and Nurlatifa, Hafzatin and Santosa, Paulus Insap},
  journal={IEEE Access},
  year={2025},
  publisher={IEEE}
}

@article{miles2024cogload,
  title={EM-COGLOAD: An investigation into age and cognitive load detection using eye tracking and deep learning},
  author={Miles, Gabriella and Smith, Melvyn and Zook, Nancy and Zhang, Wenhao},
  journal={Computational and Structural Biotechnology Journal},
  volume={24},
  pages={264--280},
  year={2024},
  publisher={Elsevier}
}

@article{binz2023using,
  title={Using cognitive psychology to understand GPT-3},
  author={Binz, Marcel and Schulz, Eric},
  journal={Proceedings of the National Academy of Sciences},
  volume={120},
  number={6},
  pages={e2218523120},
  year={2023},
  publisher={National Academy of Sciences}
}

@article{brown2020language,
  title={Language models are few-shot learners},
  author={Brown, Tom and Mann, Benjamin and Ryder, Nick and Subbiah, Melanie and Kaplan, Jared D and Dhariwal, Prafulla and Neelakantan, Arvind and Shyam, Pranav and Sastry, Girish and Askell, Amanda and others},
  journal={Advances in neural information processing systems},
  volume={33},
  pages={1877--1901},
  year={2020}
}

@inproceedings{salemi2024lamp,
  title={Lamp: When large language models meet personalization},
  author={Salemi, Alireza and Mysore, Sheshera and Bendersky, Michael and Zamani, Hamed},
  booktitle={Proceedings of the 62nd Annual Meeting of the Association for Computational Linguistics (Volume 1: Long Papers)},
  pages={7370--7392},
  year={2024}
}

@article{bhatti2025clare,
  title={Clare: Cognitive load assessment in real-time with multimodal data},
  author={Bhatti, Anubhav and Angkan, Prithila and Behinaein, Behnam and Mahmud, Zunayed and Rodenburg, Dirk and Braund, Heather and Mclellan, P James and Ruberto, Aaron and Harrison, Geoffery and Wilson, Daryl and others},
  journal={IEEE Transactions on Cognitive and Developmental Systems},
  year={2025},
  publisher={IEEE}
}

@inproceedings{skaramagkas2021cognitive,
  title={Cognitive workload level estimation based on eye tracking: A machine learning approach},
  author={Skaramagkas, Vasileios and Ktistakis, Emmanouil and Manousos, Dimitris and Tachos, Nikolaos S and Kazantzaki, Eleni and Tripoliti, Evanthia E and Fotiadis, Dimitrios I and Tsiknakis, Manolis},
  booktitle={2021 IEEE 21st International Conference on Bioinformatics and Bioengineering (BIBE)},
  pages={1--5},
  year={2021},
  organization={IEEE}
}

@article{chen2025machine,
  title={Machine Learning Models to Predict Individual Cognitive Load in Collaborative Learning: Combining fNIRS and Eye-Tracking Data},
  author={Chen, Wenli and Lin, Zirou and Zheng, Lishan and Ho, Mei-Yee Mavis and Ali, Farhan and Teo, Wei Peng},
  journal={Machine Learning and Knowledge Extraction},
  volume={7},
  number={2},
  pages={51},
  year={2025},
  publisher={MDPI}
}

@inproceedings{nishitha2023eye,
  title={Eye-cog: Eye tracking-based deep learning model for the detection of cognitive impairments in college students},
  author={Nishitha, U and Kandimalla, Revanth and Jyotsna, C and Singh, Tripty and others},
  booktitle={2023 14th International Conference on Computing Communication and Networking Technologies (ICCCNT)},
  pages={1--7},
  year={2023},
  organization={IEEE}
}

@article{rahman2021vision,
  title={Vision-based driver’s cognitive load classification considering eye movement using machine learning and deep learning},
  author={Rahman, Hamidur and Ahmed, Mobyen Uddin and Barua, Shaibal and Funk, Peter and Begum, Shahina},
  journal={Sensors},
  volume={21},
  number={23},
  pages={8019},
  year={2021},
  publisher={MDPI}
}

@article{lewis2020retrieval,
  title={Retrieval-augmented generation for knowledge-intensive nlp tasks},
  author={Lewis, Patrick and Perez, Ethan and Piktus, Aleksandra and Petroni, Fabio and Karpukhin, Vladimir and Goyal, Naman and K{\"u}ttler, Heinrich and Lewis, Mike and Yih, Wen-tau and Rockt{\"a}schel, Tim and others},
  journal={Advances in neural information processing systems},
  volume={33},
  pages={9459--9474},
  year={2020}
}

@article{wu2024survey,
  title={A survey on large language models for recommendation},
  author={Wu, Likang and Zheng, Zhi and Qiu, Zhaopeng and Wang, Hao and Gu, Hongchao and Shen, Tingjia and Qin, Chuan and Zhu, Chen and Zhu, Hengshu and Liu, Qi and others},
  journal={World Wide Web},
  volume={27},
  number={5},
  pages={60},
  year={2024},
  publisher={Springer}
}

@article{wei2022chain,
  title={Chain-of-thought prompting elicits reasoning in large language models},
  author={Wei, Jason and Wang, Xuezhi and Schuurmans, Dale and Bosma, Maarten and Xia, Fei and Chi, Ed and Le, Quoc V and Zhou, Denny and others},
  journal={Advances in neural information processing systems},
  volume={35},
  pages={24824--24837},
  year={2022}
}

@inproceedings{dong2024survey,
  title={A survey on in-context learning},
  author={Dong, Qingxiu and Li, Lei and Dai, Damai and Zheng, Ce and Ma, Jingyuan and Li, Rui and Xia, Heming and Xu, Jingjing and Wu, Zhiyong and Chang, Baobao and others},
  booktitle={Proceedings of the 2024 conference on empirical methods in natural language processing},
  pages={1107--1128},
  year={2024}
}

@article{achiam2023gpt,
  title={Gpt-4 technical report},
  author={Achiam, Josh and Adler, Steven and Agarwal, Sandhini and Ahmad, Lama and Akkaya, Ilge and Aleman, Florencia Leoni and Almeida, Diogo and Altenschmidt, Janko and Altman, Sam and Anadkat, Shyamal and others},
  journal={arXiv preprint arXiv:2303.08774},
  year={2023}
}

@article{steichen2014inferring,
  title={Inferring visualization task properties, user performance, and user cognitive abilities from eye gaze data},
  author={Steichen, Ben and Conati, Cristina and Carenini, Giuseppe},
  journal={ACM Transactions on Interactive Intelligent Systems (TiiS)},
  volume={4},
  number={2},
  pages={1--29},
  year={2014},
  publisher={ACM New York, NY, USA}
}

@book{holmqvist2011eye,
  title={Eye tracking: A comprehensive guide to methods and measures},
  author={Holmqvist, Kenneth and Nystr{\"o}m, Marcus and Andersson, Richard and Dewhurst, Richard and Jarodzka, Halszka and Van de Weijer, Joost},
  year={2011},
  publisher={oup Oxford}
}

@inproceedings{borgeaud2022improving,
  title={Improving language models by retrieving from trillions of tokens},
  author={Borgeaud, Sebastian and Mensch, Arthur and Hoffmann, Jordan and Cai, Trevor and Rutherford, Eliza and Millican, Katie and Van Den Driessche, George Bm and Lespiau, Jean-Baptiste and Damoc, Bogdan and Clark, Aidan and others},
  booktitle={International conference on machine learning},
  pages={2206--2240},
  year={2022},
  organization={PMLR}
}

@article{chiossi2024understanding,
  title={Understanding the impact of the reality-virtuality continuum on visual search using fixation-related potentials and eye tracking features},
  author={Chiossi, Francesco and Gruenefeld, Uwe and Hou, Baosheng James and Newn, Joshua and Ou, Changkun and Liao, Rulu and Welsch, Robin and Mayer, Sven},
  journal={Proceedings of the ACM on Human-Computer Interaction},
  volume={8},
  number={MHCI},
  pages={1--33},
  year={2024},
  publisher={ACM New York, NY, USA}
}

@article{pillai2020response,
  title={Response time and eye tracking datasets for activities demanding varying cognitive load},
  author={Pillai, Prarthana and Ayare, Prathamesh and Balasingam, Balakumar and Milne, Kevin and Biondi, Francesco},
  journal={Data in brief},
  volume={33},
  pages={106389},
  year={2020},
  publisher={Elsevier}
}

@article{angkan2024multimodal,
  title={Multimodal brain--computer interface for in-vehicle driver cognitive load measurement: Dataset and baselines},
  author={Angkan, Prithila and Behinaein, Behnam and Mahmud, Zunayed and Bhatti, Anubhav and Rodenburg, Dirk and Hungler, Paul and Etemad, Ali},
  journal={IEEE Transactions on Intelligent Transportation Systems},
  volume={25},
  number={6},
  pages={5949--5964},
  year={2024},
  publisher={IEEE}
}

@inproceedings{radford2023robust,
  title={Robust speech recognition via large-scale weak supervision},
  author={Radford, Alec and Kim, Jong Wook and Xu, Tao and Brockman, Greg and McLeavey, Christine and Sutskever, Ilya},
  booktitle={International conference on machine learning},
  pages={28492--28518},
  year={2023},
  organization={PMLR}
}

@article{grattafiori2024llama,
  title={The llama 3 herd of models},
  author={Grattafiori, Aaron and Dubey, Abhimanyu and Jauhri, Abhinav and Pandey, Abhinav and Kadian, Abhishek and Al-Dahle, Ahmad and Letman, Aiesha and Mathur, Akhil and Schelten, Alan and Vaughan, Alex and others},
  journal={arXiv preprint arXiv:2407.21783},
  year={2024}
}

@article{sendhilnathan2024implicit,
  title={Implicit gaze research for XR systems},
  author={Sendhilnathan, Naveen and Fernandes, Ajoy S and Proulx, Michael J and Jonker, Tanya R},
  journal={arXiv preprint arXiv:2405.13878},
  year={2024}
}

@article{mansour2025enabling,
  title={Enabling eye tracking for crowd-sourced data collection with project aria},
  author={Mansour, Yusuf and Fernandes, Ajoy Savio and Somasundaram, Kiran and Hefny, Tarek and Shakeri, Mahsa and Komogortsev, Oleg and Sharma, Abhishek and Proulx, Michael J},
  journal={IEEE Access},
  year={2025},
  publisher={IEEE}
}

@inproceedings{burlingham2024real,
  title={Real-world scanpaths exhibit long-term temporal dependencies: Considerations for contextual AI for AR applications},
  author={Burlingham, Charlie S and Sendhilnathan, Naveen and Wu, Xiuyun and Murdison, T Scott and Proulx, Michael J},
  booktitle={Proceedings of the 2024 Symposium on Eye Tracking Research and Applications},
  pages={1--7},
  year={2024}
}

@inproceedings{wilson2025eye,
  title={Eye Gaze as a Signal for Conveying User Attention in Contextual AI Systems},
  author={Wilson, Ethan and Sendhilnathan, Naveen and Burlingham, Charlie S and Mansour, Yusuf and Cavin, Robert and Tetali, Sai Deep and Fernandes, Ajoy Savio and Proulx, Michael J},
  booktitle={Proceedings of the 2025 Symposium on Eye Tracking Research and Applications},
  pages={1--7},
  year={2025}
}

@inproceedings{wang2024gazesam,
  title={Gazesam: Interactive image segmentation with eye gaze and segment anything model},
  author={Wang, Bin and Aboah, Armstrong and Zhang, Zheyuan and Pan, Hongyi and Bagci, Ulas},
  booktitle={Gaze Meets Machine Learning Workshop},
  pages={254--265},
  year={2024},
  organization={PMLR}
}

@inproceedings{mondal2025gaze,
  title={Gaze-Language Alignment for Zero-Shot Prediction of Visual Search Targets from Human Gaze Scanpaths},
  author={Mondal, Sounak and Sendhilnathan, Naveen and Zhang, Ting and Liu, Yue and Proulx, Michael and Iuzzolino, Michael Louis and Qin, Chuan and Jonker, Tanya R},
  booktitle={Proceedings of the IEEE/CVF International Conference on Computer Vision},
  pages={2738--2749},
  year={2025}
}

@article{hartigan1979algorithm,
  title={Algorithm AS 136: A k-means clustering algorithm},
  author={Hartigan, John A and Wong, Manchek A},
  journal={Journal of the royal statistical society. series c (applied statistics)},
  volume={28},
  number={1},
  pages={100--108},
  year={1979},
  publisher={JSTOR}
}

@inproceedings{salvucci2000identifying,
  title={Identifying fixations and saccades in eye-tracking protocols},
  author={Salvucci, Dario D and Goldberg, Joseph H},
  booktitle={Proceedings of the 2000 symposium on Eye tracking research \& applications},
  pages={71--78},
  year={2000}
}

@article{paas1992training,
  title={Training strategies for attaining transfer of problem-solving skill in statistics: a cognitive-load approach.},
  author={Paas, Fred GWC},
  journal={Journal of educational psychology},
  volume={84},
  number={4},
  pages={429},
  year={1992},
  publisher={American Psychological Association}
}

@article{van2012timing,
  title={Timing and frequency of mental effort measurement: Evidence in favour of repeated measures},
  author={Van Gog, Tamara and Kirschner, Femke and Kester, Liesbeth and Paas, Fred},
  journal={Applied cognitive psychology},
  volume={26},
  number={6},
  pages={833--839},
  year={2012},
  publisher={Wiley Online Library}
}

@book{dunteman1989principal,
  title={Principal components analysis},
  author={Dunteman, George H},
  volume={69},
  year={1989},
  publisher={Sage}
}

@inproceedings{cury2019hybrid,
  title={Hybrid methodology: combining ethnography, cognitive science, and machine learning to inform the development of context-aware personal computing and assistive technology},
  author={Cury, Maria and Whitworth, Eryn and Barfort, Sebastian and Bochereau, S{\'e}r{\'e}na and Browder, Jonathan and Jonker, Tanya R and Kim, Kahyun Sophie and Krenchel, Mikkel and RAMSEY-ELLIOT, MORGAN and Sch{\"u}{\"u}r, Friederike and others},
  booktitle={Ethnographic praxis in industry conference proceedings},
  volume={2019},
  number={1},
  pages={254--281},
  year={2019},
  organization={Wiley Online Library}
}

@article{jin2023time,
  title={Time-llm: Time series forecasting by reprogramming large language models},
  author={Jin, Ming and Wang, Shiyu and Ma, Lintao and Chu, Zhixuan and Zhang, James Y and Shi, Xiaoming and Chen, Pin-Yu and Liang, Yuxuan and Li, Yuan-Fang and Pan, Shirui and others},
  journal={arXiv preprint arXiv:2310.01728},
  year={2023}
}

@article{van2018pupil,
  title={Pupil dilation as an index of effort in cognitive control tasks: A review},
  author={Van der Wel, Pauline and Van Steenbergen, Henk},
  journal={Psychonomic bulletin \& review},
  volume={25},
  number={6},
  pages={2005--2015},
  year={2018},
  publisher={Springer}
}

@article{rayner1998eye,
  title={Eye movements in reading and information processing: 20 years of research.},
  author={Rayner, Keith},
  journal={Psychological bulletin},
  volume={124},
  number={3},
  pages={372},
  year={1998},
  publisher={American Psychological Association}
}

@article{bachurina2022multiple,
  title={Multiple levels of mental attentional demand modulate peak saccade velocity and blink rate},
  author={Bachurina, Valentina and Arsalidou, Marie},
  journal={Heliyon},
  volume={8},
  number={1},
  year={2022},
  publisher={Elsevier}
}

@article{ekin2025prediction,
  title={Prediction of intrinsic and extraneous cognitive load with oculometric and biometric indicators},
  author={Ekin, Merve and Krejtz, Krzysztof and Duarte, Carlos and Duchowski, Andrew T and Krejtz, Izabela},
  journal={Scientific Reports},
  volume={15},
  number={1},
  pages={5213},
  year={2025},
  publisher={Nature Publishing Group UK London}
}

@article{sevcenko2023theory,
  title={Theory-based approach for assessing cognitive load during time-critical resource-managing human--computer interactions: an eye-tracking study},
  author={Sevcenko, Natalia and Appel, Tobias and Ninaus, Manuel and Moeller, Korbinian and Gerjets, Peter},
  journal={Journal on Multimodal User Interfaces},
  volume={17},
  number={1},
  pages={1--19},
  year={2023},
  publisher={Springer}
}

@article{wilson2024privacy,
  title={Privacy-preserving gaze data streaming in immersive interactive virtual reality: Robustness and user experience},
  author={Wilson, Ethan and Ibragimov, Azim and Proulx, Michael J and Tetali, Sai Deep and Butler, Kevin and Jain, Eakta},
  journal={IEEE Transactions on Visualization and Computer Graphics},
  volume={30},
  number={5},
  pages={2257--2268},
  year={2024},
  publisher={IEEE}
}

@article{raju2025real,
  title={Real-Time Lightweight Gaze Privacy-Preservation Techniques Validated via Offline Gaze-Based Interaction Simulation},
  author={Raju, Mehedi Hasan and Komogortsev, Oleg V},
  journal={arXiv preprint arXiv:2511.09846},
  year={2025}
}

@article{wickens2008multiple,
  title={Multiple resources and mental workload},
  author={Wickens, Christopher D},
  journal={Human factors},
  volume={50},
  number={3},
  pages={449--455},
  year={2008},
  publisher={SAGE Publications Sage CA: Los Angeles, CA}
}

\newpage
\appendix

\section*{Appendix}
The appendix of this paper includes:

Appendix \ref{A_source}: Analysis of source of performance gain, fairness comparison with LLM and non-LLM methods

Appendix \ref{C}: Latency, Robustness, Interpretability, Label Reliability Analysis
\begin{itemize}
    \item \ref{A_time} Inference latency analysis 
    \item \ref{A_Robustness} Model robustness analysis
    \item \ref{A_Interpretability} Model interpretability and faithfulness analysis
    \item \ref{A_label} Label reliability analysis
    \item \ref{A_userbeh} Correlation analysis between user profile and behavioral outcomes
\end{itemize}

Appendix \ref{B}: Additional Experiments
\begin{itemize}
    \item \ref{A_cross} Cross-LLM generalization
    \item \ref{taskdesign} Reasons of our task design
    \item \ref{A_crosstask} Cross-Task transfer and generalization ability
    \item \ref{A_ablationtemp} Ablation study on temporal window length
    \item \ref{A_task} Performance for each task scenario
\end{itemize}

Appendix \ref{A}: Implementation Details
\begin{itemize}
    \item \ref{A_implementation} Implementation details for comparison methods
    \item \ref{A_features} Definition and calculation details of gaze features and raw gaze processing
    \item \ref{A_user} User profiling and clustering
    \item \ref{A_prompt} Prompt design
\end{itemize}

Appendix \ref{A_data}: More details of our collected dataset CogLoad-Bench

Appendix \ref{A_privacy}: Impact statement (ethical and privacy discussion)

\newpage
\section{Analysis of source of performance gain, fairness comparison with LLM and non-LLM methods} \label{A_source}

A natural question is whether GazeMind's substantial improvement comes from (1) the additional context provided to the LLM, (2) the LLM's reasoning capability, or (3) the combination enabled by our structured module design. To isolate these factors, we conduct two controlled comparisons that decouple LLM reasoning from context learning.

\paragraph{Non-LLM model with the same context.}
We provide a Transformer with the exact same context as GazeMind, encoded numerically: TGR task-specific rules are converted into one-hot task indicators with feature importance ranks and direction signs, AUP user profiles are encoded as one-hot profile types augmented with personal baselines, and CogRAG retrieved examples (using identical similarity metric and retrieval pipeline) are flattened into feature tables paired with their labels. This setup gives the Transformer access to all information available to GazeMind. As shown in Table~\ref{tab:transformer_context}, the additional context substantially improves the Transformer (37.16\% $\rightarrow$ 45.38\%), confirming that the context is informative. However, performance remains far below GazeMind (62.73\%), indicating that \textbf{extra context alone is insufficient without LLM reasoning to interpret it}.

\begin{table}[h]
\centering
\small
\caption{Performance of a non-LLM model (Transformer) given the same context as GazeMind.}
\label{tab:transformer_context}
\begin{tabular}{lcccc}
\toprule
\textbf{Method} & \textbf{Accuracy $\uparrow$} & \textbf{Precision $\uparrow$} & \textbf{Recall $\uparrow$} & \textbf{F1 $\uparrow$} \\
\midrule
Transformer            & 37.16 & 34.55 & 34.40 & 33.15 \\
Transformer + context  & 45.38 & 44.43 & 44.49 & 44.43 \\
\midrule
\textbf{GazeMind (LLM + context)} & \textbf{62.73} & \textbf{62.58} & \textbf{61.77} & \textbf{62.11} \\
\bottomrule
\end{tabular}
\end{table}

\paragraph{LLM with the same context but conventional ICL.}
We further evaluate whether the gain comes from naively giving the LLM more context, or specifically from how our modules structure the context. We construct a \emph{Naive ICL} variant where GPT-4o receives the same context access as GazeMind but through conventional approaches instead of our proposed modules. Specifically, TGR's LLM-generated guidance rules are replaced with Decision Tree rules derived from the same statistics, AUP's user profile traits are replaced with raw per-user pupil baselines (pupil size, blink count, gaze stability, pupil sensitivity) and population-level means, and CogRAG's temporal-similarity retrieval is replaced with standard vector-distance retrieval over flattened feature vectors to retrieve three similar samples with labels. As shown in Table~\ref{tab:naive_icl}, GPT-4o + Naive ICL achieves only 44.19\% accuracy, substantially below GazeMind despite identical context access. This confirms that \textbf{the performance advantage comes from how the context is structured and presented to the LLM, not from context access itself}.

\begin{table}[h]
\centering
\small
\caption{Performance of GPT-4o given the same data access as GazeMind but through conventional ICL.}
\label{tab:naive_icl}
\begin{tabular}{lcccc}
\toprule
\textbf{Method} & \textbf{Accuracy $\uparrow$} & \textbf{Precision $\uparrow$} & \textbf{Recall $\uparrow$} & \textbf{F1 $\uparrow$} \\
\midrule
GPT-4o (zero-shot)        & 39.62 & 40.81 & 40.61 & 38.98 \\
GPT-4o + Naive ICL        & 44.19 & 42.17 & 41.83 & 41.81 \\
\midrule
\textbf{GazeMind} (GPT-4o + TGR + AUP + CogRAG) & \textbf{62.73} & \textbf{62.58} & \textbf{61.77} & \textbf{62.11} \\
\bottomrule
\end{tabular}
\end{table}

\paragraph{Conclusion.}
Together, these two experiments form a complementary pair: Table~\ref{tab:transformer_context} shows that extra context alone is not sufficient without LLM reasoning, while Table~\ref{tab:naive_icl} shows that LLM reasoning alone is not sufficient without proper context representation. Neither extra context alone nor LLM reasoning with conventional context structuring approaches GazeMind's performance. This confirms that \textbf{GazeMind's performance gain comes from LLM reasoning combined with context design by our proposed module, not from extra context or LLM reasoning alone.}


\newpage
\section{Latency, Robustness, Interpretability, Label Reliability Analysis} \label{C}

\subsection{Inference Latency Analysis} \label{A_time}
Deploying cognitive load assessment on smart glasses requires low inference latency. Importantly, GazeMind does not require second-by-second prediction. Instead, it predicts cognitive load once per time window and back-propagates the predicted label to the preceding interval. In our experiments, each inference takes approximately 2 seconds, which is well within the 5-second minimum time window used in this study. Note that this latency includes the API calls to GPT-4o, while on-device deployed models would achieve even lower latency. 

To further evaluate on-device deployment efficiency, we replace GPT-4o with Qwen3.5-4B, a small open LLM. As shown in Table \ref{tab:small_llm}, Qwen3.5-4B achieves 50.62\% accuracy at 0.45s latency, still well above all baselines (33-39\%, Table \ref{tab:main_results}) while being nearly 5x faster than GPT-4o's cloud API.

\begin{table}[h]
\centering
\caption{Performance and inference latency of GazeMind with different LLM backbones. Qwen3.5-4B achieves $\sim$5$\times$ faster inference than GPT-4o while still outperforming all baselines in Table~\ref{tab:main_results}, demonstrating feasibility for on-device deployment.}
\label{tab:small_llm}
\begin{tabular}{lccccc}
\toprule
\textbf{LLM Backbone} & \textbf{Accuracy $\uparrow$} & \textbf{Precision $\uparrow$} & \textbf{Recall $\uparrow$} & \textbf{F1 $\uparrow$} & \textbf{Latency (s) $\downarrow$} \\
\midrule
Qwen3.5-4B & 50.62 & 57.60 & 56.68 & 48.00 & \textbf{0.45} \\
GPT-4o     & \textbf{62.73} & \textbf{62.58} & \textbf{61.77} & \textbf{62.11} & 2.15 \\
\bottomrule
\end{tabular}
\end{table}

We still acknowledge that further optimization is required and possible. But overall, the current latency is sufficient for practical deployment under our inference strategy.

\subsection{Model Robustness Analysis} \label{A_Robustness}

We analyze GazeMind's robustness from two perspectives that reflect realistic deployment conditions: (1) noise in self-reported cognitive load labels, and (2) missing values in raw gaze data due to tracking loss or sensor failures.

\subsubsection{Robustness to Label Noise}

Self-reported cognitive load labels inevitably carry noise from boundary ambiguity between adjacent levels and inter-participant variability. We further use temporal interpolation to back-propagate verbal reports into per-second labels, which may introduce additional noise. We evaluate GazeMind's robustness against following two noise sources.

\paragraph{Effect of label interpolation.}
We evaluate GazeMind on only the timestamps with reported labels, removing all interpolated labels. As shown in Table~\ref{tab:label_noise}, evaluating without interpolation yields slightly higher accuracy (65.47\% vs.\ 62.73\%), indicating that interpolation introduces modest noise ($\sim$3 points). This is a worthwhile trade-off, as interpolation provides dense temporal labels essential for database construction and time-series modeling.

\paragraph{Effect of label noise.}
We further inject realistic label noise by randomly flipping a fraction of labels to adjacent levels (low~$\leftrightarrow$~moderate, moderate~$\leftrightarrow$~high). At 5\%, 10\%, and 20\% noise rates, accuracy gradually decreases but remains substantially above all baselines in Table~\ref{tab:main_results}, exceeding the best baseline by $\sim$15\% even at 20\% noise. This robustness comes from three design choices: TGR rules are derived from population-level statistics rather than individual labels, AUP does not use labels at all, and CogRAG's validation mechanism allows the LLM to reject inconsistent retrieved examples. Together, these reduce the impact of noisy labels on the framework.

\begin{table}[h]
\centering
\small
\caption{Robustness of GazeMind to label noise. We compare full interpolated labels, only reported timestamps (no interpolation), and added label noise at 5\%, 10\%, and 20\%.}
\label{tab:label_noise}
\begin{tabular}{lcccc}
\toprule
\textbf{Setting} & \textbf{Accuracy $\uparrow$} & \textbf{Precision $\uparrow$} & \textbf{Recall $\uparrow$} & \textbf{F1 $\uparrow$} \\
\midrule
GazeMind (full interpolation)  & 62.73 & 62.58 & 61.77 & 62.11 \\
Only reported timestamps       & 65.47 & 65.30 & 64.18 & 64.62 \\
\midrule
+ 5\% label noise              & 60.39 & 60.35 & 58.97 & 59.45 \\
+ 10\% label noise             & 58.41 & 58.39 & 57.20 & 57.54 \\
+ 20\% label noise             & 54.22 & 53.90 & 53.26 & 53.28 \\
\bottomrule
\end{tabular}
\end{table}

\subsubsection{Robustness to Missing Gaze Data}

In real-world smart glasses deployment, raw gaze signals may be lost due to tracking failures, blinks, or off-screen gaze. We conduct a stress test where a fraction of raw gaze samples is randomly dropped and filled by interpolation before feature extraction. Table~\ref{tab:missing_gaze} shows GazeMind's accuracy at 0\%, 10\%, 30\%, and 50\% missing ratios. At 10\% missing, accuracy slightly improves over the no-missing baseline, likely because random missing samples coincide with noisy gaze points that interpolation effectively smooths. Even at 50\% missing, accuracy remains above all baselines in Table~\ref{tab:main_results}, demonstrating the framework's robustness to real-world data quality variations.

\begin{table}[h]
\centering
\small
\caption{Robustness of GazeMind to missing values in raw gaze data.}
\label{tab:missing_gaze}
\begin{tabular}{cc}
\toprule
\textbf{Missing Ratio (\%)} & \textbf{Accuracy (\%) $\uparrow$} \\
\midrule
0   & 62.73 \\
10  & \textbf{64.56} \\
30  & 56.94 \\
50  & 47.54 \\
\bottomrule
\end{tabular}
\end{table}

\subsubsection{Robustness to Subtask Variation} 
Although task instructions are pre-specified, real tasks naturally contain subtask variation (e.g., card-playing vs.\ waiting in the social game). GazeMind handles this through two mechanisms: (1) it predicts cognitive load per 5-second window rather than per task, capturing within-task fluctuations across different subtasks, (2) CogRAG retrieves reference samples based on gaze pattern similarity within the same task category, implicitly retrieving samples whose gaze patterns match the current subtask. Empirically, the Gaming task contains more diverse subtasks than Reading, yet GazeMind achieves comparable performance on both (Table~\ref{tab:task_results}), demonstrating robustness to subtask variation.

\subsection{Model Interpretability and Faithfulness Analysis} \label{A_Interpretability}

A central claim of GazeMind is that it provides interpretable predictions grounded in genuine cognitive reasoning rather than plausible-sounding but spurious explanations. To validate this, we conduct two complementary analyses: (1) verifying that TGR-generated rules align with established cognitive science literature, and (2) evaluating interpretability of GazeMind by a counterfactual perturbation test.

\subsubsection{TGR Rule Alignment with Cognitive Science Literature}

The TGR module automatically generates task-specific interpretation rules from population-level feature statistics. A key concern is whether these rules reflect genuine cognitive signals or capture dataset-specific spurious correlations. To address this, we cross-reference all feature-task pairs in the TGR rules (ranked by feature importance) against established findings in the cognitive science literature. Note that TGR retains only features that exhibit discriminative distributions across cognitive load levels, features with substantially overlapping distributions are excluded from the rule. Task-universal features (e.g., pupil\_size) share references across tasks, while task-dependent features (e.g., saccade\_ratio) cite task-specific studies.

Table~\ref{tab:tgr_literature} lists all feature-task pairs, the directional relationship between each feature and cognitive load identified by TGR, and supporting references from cognitive science literature.

\begin{table}[h]
\centering
\small
\caption{TGR rule alignment with cognitive science literature. For each task, we list features ranked by TGR importance, the directional relationship to cognitive load, and supporting references.}
\label{tab:tgr_literature}
\resizebox{\linewidth}{!}{
\begin{tabular}{llcl}
\toprule
\textbf{Task} & \textbf{Feature (Rank)} & \textbf{Direction to higher cognitive load} & \textbf{References} \\
\midrule
Reading & pupil\_size (\#1)         & $\uparrow$   & \cite{van2018pupil, kosch2023survey} \\
        & fixation\_duration (\#2)  & $\uparrow$   & \cite{rayner1998eye, kosch2023survey} \\
        & saccade\_ratio (\#3)      & $\downarrow$ & \cite{rayner1998eye, kosch2023survey} \\
        & blink\_count (\#4)        & $\downarrow$ & \cite{kosch2023survey, bachurina2022multiple} \\
        & saccade\_duration (\#5)   & $\downarrow$ & \cite{rayner1998eye, kosch2023survey} \\
\midrule
Gaming  & saccade\_ratio (\#1)      & $\uparrow$   & \cite{ekin2025prediction, sevcenko2023theory} \\
        & blink\_count (\#2)        & $\downarrow$ & \cite{kosch2023survey} \\
        & pupil\_size (\#3)         & $\uparrow$   & \cite{van2018pupil, kosch2023survey} \\
        & saccade\_amplitude (\#4)  & $\downarrow$ & \cite{ekin2025prediction,rayner1998eye} \\
\bottomrule
\end{tabular}}
\end{table}

\textbf{All directional relationships in TGR are consistent with established findings in the cognitive science literature.} Notably, saccade\_ratio show opposite trends across two tasks, reflecting tasks' distinct gaze behavior. Reading demands sustained focus, so higher load increases fixation and suppresses saccade \cite{rayner1998eye,kosch2023survey}. In contrast, Gaming requires players to constantly shift gaze among cards and opponents, so higher load elevates saccade ratio \cite{ekin2025prediction,sevcenko2023theory}. These patterns are well documented in prior studies under corresponding tasks and are precisely captured by TGR. Hence, \textbf{the convergence of TGR rules with established findings confirms genuine cognitive signal rather than spurious correlation.}

\subsubsection{Counterfactual Perturbation Analysis}

We further evaluate whether GazeMind's predictions and explanations are faithfully grounded in input gaze features through a counterfactual perturbation test. The core idea is to verify that prediction changes and explanations vary predictably with controlled feature manipulations, rather than being post-hoc rationalizations.

\paragraph{Experimental setup.}
We randomly select 100 test samples from the Reading task and another 100 from the Gaming task. For each task, we perturb the top-3 features ranked by TGR importance one at a time. For each sample, we flip the z-score sign of the target feature across all timesteps in the feature table while keeping all other features unchanged, then run GazeMind and a GPT-4o baseline (with TGE-encoded gaze features but no other modules) on both the original and perturbed inputs. We compare predictions and explanations between the two conditions across three metrics:
\begin{itemize}
\item \textbf{Flip Rate}: the percentage of samples whose prediction changes after perturbation. A higher flip rate indicates predictions are more sensitive to perturbed features.
\item \textbf{Direction Correct}: the percentage of explanations that correctly describe the perturbed feature's direction (e.g., correctly identifying that the perturbed pupil\_size now shows a decrease).
\item \textbf{Attribution Correct}: among samples whose predictions flipped, the percentage that correctly attribute the flip to the perturbed feature in the explanation.
\end{itemize}

Table~\ref{tab:counterfactual} reports counterfactual results for both GazeMind and GPT-4o across all six feature-task combinations.

\begin{table}[h]
\centering
\small
\caption{Counterfactual perturbation analysis on top-3 TGR-ranked features per task, comparing GazeMind and GPT-4o (with TGE only). For each feature, we report Flip Rate, Direction Correct, and Attribution Correct based on perturbations of 100 randomly selected test samples per task.}
\label{tab:counterfactual}
\resizebox{\linewidth}{!}{
\begin{tabular}{llcccc}
\toprule
\textbf{Model} & \textbf{Task} & \textbf{Perturbed Feature (Rank)} & \textbf{Flip Rate $\uparrow$} & \textbf{Direction Correct $\uparrow$} & \textbf{Attribution Correct $\uparrow$} \\
\midrule
\multirow{6}{*}{GazeMind}
 & Reading & pupil\_size (\#1)         & 61\% & 100\% & 97.4\% \\
 & Reading & fixation\_duration (\#2)  & 48\% & 100\% & 87.0\% \\
 & Reading & saccade\_ratio (\#3)      & 30\% & 94\%  & 93.3\% \\
 & Gaming  & saccade\_ratio (\#1)      & 48\% & 100\% & 91.8\% \\
 & Gaming  & blink\_count (\#2)        & 31\% & 100\% & 92.0\% \\
 & Gaming  & pupil\_size (\#3)         & 24\% & 95\%  & 85.7\% \\
\midrule
\multirow{6}{*}{GPT-4o}
 & Reading & pupil\_size (\#1)         & 10\% & 93\%  & 46.7\% \\
 & Reading & fixation\_duration (\#2)  & 17\% & 92\%  & 34.8\% \\
 & Reading & saccade\_ratio (\#3)      & 7\%  & 100\% & 31.8\% \\
 & Gaming  & saccade\_ratio (\#1)      & 15\% & 93\%  & 60.0\% \\
 & Gaming  & blink\_count (\#2)        & 16\% & 95\%  & 26.2\% \\
 & Gaming  & pupil\_size (\#3)         & 23\% & 99\%  & 28.6\% \\
\bottomrule
\end{tabular}}
\end{table}

We highlight three findings:

\paragraph{Finding 1: Explanation sensitivity tracks TGR feature importance for GazeMind but not for GPT-4o.}
For GazeMind, flip rate monotonically decreases with TGR feature importance rank in both tasks (Reading: 61\% $\rightarrow$ 48\% $\rightarrow$ 30\%; Gaming: 48\% $\rightarrow$ 31\% $\rightarrow$ 24\%), indicating that perturbing more important features causes more prediction changes, as expected. In contrast, GPT-4o shows no such monotonic pattern (Reading: 10\% $\rightarrow$ 17\% $\rightarrow$ 7\%; Gaming: 15\% $\rightarrow$ 16\% $\rightarrow$ 23\%). Moreover, GazeMind's consistently higher flip rates indicate that its predictions are genuinely grounded in gaze features, whereas GPT-4o's low and unstructured flip rates suggest that without TGR guidance, the LLM lacks systematic reasoning over gaze features.

\paragraph{Finding 2: Both models can describe perturbed feature directions.}
Direction correctness is high for both models ($>$92\% in nearly all cases). This is expected, as LLMs are generally capable of describing input feature values. However, accurate description does not imply accurate interpretation, especially in this unfamiliar physiological reasoning task, where the model needs additional context and guidance to correctly relate the described values to cognitive load.

\paragraph{Finding 3: Attribution correctness is much higher for GazeMind than GPT-4o.}
GazeMind reliably traces prediction flips back to the specific perturbed feature (85.7--97.4\%), while GPT-4o frequently attributes flips to the wrong feature (26.2--60.0\%). This shows that the ability to reason \emph{faithfully} over gaze features comes from GazeMind's structured modules, not from inherent LLM capabilities. Among GazeMind's unchanged-prediction cases, most explanations (around 90\%) explicitly acknowledge the perturbed feature change but provide justified reasoning based on other evidence sources (other features, user profile, CogRAG examples), demonstrating genuine multi-source reasoning rather than single-feature dependency or fabricated explanations. This shows that the ability to reason faithfully over gaze features comes from GazeMind's structured modules, not from the LLM itself.

Together, the literature alignment of TGR rules and the counterfactual perturbation analysis confirm that GazeMind's interpretability is grounded in genuine cognitive reasoning rather than plausible-sounding but spurious explanations.

\subsection{Label Reliability Analysis} \label{A_label}

A natural concern with self-reported cognitive load labels is whether they reflect genuine cognitive signals or arbitrary subjective ratings. Since the entire framework is based on these labels, understanding their reliability is important for interpreting the $\sim$60\% accuracy ceiling. We validate label reliability from three complementary perspectives: alignment with objective task difficulty, intra-participant consistency, and inter-participant agreement.

\paragraph{Task difficulty alignment.}
We use the manipulated task difficulty (easy, medium, hard) as an \textbf{objective reference} and analyze how well self-reported labels align with it. Across all participants, self-reported labels align well with objective task difficulty with a Spearman correlation of $\rho = 0.45$ ($p < 0.001$), comparable to typical correlations reported in cognitive load literature. Furthermore, across the three difficulty levels, the matching cognitive load category is the most frequently reported label: 71.0\% of Easy trials are reported as Low, 43.8\% of Medium trials as Moderate, and 45.3\% of Hard trials as High. This consistent alignment with objective difficulty indicates that \textbf{the labels carry genuine cognitive signal rather than random noise.}

\paragraph{Intra-participant consistency.}
We analyze within-subject label variance across trials of the same difficulty level for each participant. Across all participants, within-subject variance is low (Easy: 0.174, Medium: 0.313, Hard: 0.312), indicating that \textbf{each participant uses the rating scale consistently across trials of the same difficulty}, especially under clear cognitive demands. The slightly higher variance under Medium and Hard conditions reflects greater individual variability when task demand is more cognitively challenging.

\paragraph{Inter-participant agreement.}
We further analyze how many participants agree on the same label for trials of the same difficulty. Inter-participant agreement reaches 80.5\% on Easy trials reported as Low, 53.0\% on Medium trials reported as Moderate, and 49.7\% on Hard trials reported as High, yielding an average agreement of $\sim$61\% across all difficulty levels. Notably, this $\sim$61\% inter-participant agreement closely matches GazeMind's accuracy of 62.73\%, suggesting that the $\sim$60\% accuracy ceiling reflects the inherent subjectivity of cognitive load reporting rather than a fundamental limitation of the model.

Together, these three analyses confirm that the self-reported labels carry genuine cognitive signal aligned with objective task difficulty, exhibit consistent within-subject usage, and converge across participants at a rate that closely matches GazeMind's classification accuracy.

\subsection{Correlation Analysis between User Profile and Behavioral Outcomes} \label{A_userbeh}

A potential concern with the AUP module is whether the three K-Means clusters (High-Reactor, Low-Reactor, Restless) reflect statistically separable groups in feature space instead of psychologically meaningful individual differences. To validate that our profiles capture genuine behavioral characteristics, we analyze how each profile correlates with behavioral outcomes computed independently of the clustering features. Specifically, we compute the following four behavioral measures for each profile group:

\begin{itemize}
\item \textbf{Age}: average age of participants in each profile group.
\item \textbf{Hard Task Sensitivity}: the percentage of High cognitive load reports during Hard task trials, reflecting how strongly each profile perceives high-demand conditions.
\item \textbf{Perception Bias}: the average difference between reported cognitive load and objective task difficulty, where positive values indicate overestimation and negative values indicate underestimation.
\item \textbf{Fatigue Effect}: the change in reported cognitive load over the course of a session, where positive values indicate increasing reported load over time.
\end{itemize}

Note that none of these behavioral measures use the four gaze features (blink intensity, pupil sensitivity, pupil size baseline, gaze instability) that drive the K-Means clustering, ensuring that any observed differences are independent of the clustering procedure itself.

\begin{table}[h]
\centering
\small
\caption{Behavioral outcomes across the three user profiles.}
\label{tab:user_behavioral}
\begin{tabular}{lccc}
\toprule
\textbf{Behavioral Measure} & \textbf{Low-Reactor} & \textbf{High-Reactor} & \textbf{Restless} \\
\midrule
Age (years)                                                  & 33.1   & 28.3   & 35.6 \\
Hard Task Sensitivity (\% High load on Hard trials)          & \textbf{53\%}   & 44\%   & 38\% \\
Perception Bias (reported load minus task difficulty)          & \textbf{$+0.01$} & $-0.11$ & $-0.14$ \\
Fatigue Effect (load change over time)                       & $+0.20$ & $+0.16$ & \textbf{$-0.02$} \\
\bottomrule
\end{tabular}
\end{table}

The three profiles exhibit distinct behavioral patterns. Low-Reactors perceive actual difficulty most accurately (near-zero bias) with highest sensitivity to hard tasks, and clear fatigue over time. High-Reactors are moderately accurate but tend to slightly underestimate difficulty. Restless users underestimate task difficulty the most, are least sensitive to hard tasks, and show no fatigue effect. \textbf{These differences confirm our profiles capture meaningful individual differences.}


\section{Additional Experiment} \label{B}
\subsection{Cross-LLM Generalization} \label{A_cross}

We evaluate whether GazeMind's framework generalizes across different LLM backbones, beyond the GPT-4o used in our main experiments. We replace GPT-4o with another large open-source LLM to verify if we can have similar performance gain.

We replace GPT-4o with Llama-3.3-70B as the backbone and apply the full GazeMind framework (TGR + AUP + CogRAG) without any modification. As shown in Table~\ref{tab:llama_gain}, Llama-3.3-70B alone achieves only 39.30\% accuracy, comparable to GPT-4o under zero-shot evaluation. Adding our modules raises its accuracy to 58.23\% (+18.93\%), a comparable gain to that observed with GPT-4o (+23.11\%). This confirms that GazeMind's improvement is not specific to GPT-4o but generalizes across different LLM backbones, validating the portability of our framework.

\begin{table}[h]
\centering
\small
\caption{Performance gain of GazeMind framework with Llama-3.3-70B as backbone.}
\label{tab:llama_gain}
\begin{tabular}{lcccc}
\toprule
\textbf{Method} & \textbf{Accuracy $\uparrow$} & \textbf{Precision $\uparrow$} & \textbf{Recall $\uparrow$} & \textbf{F1 $\uparrow$} \\
\midrule
Llama-3.3-70B (zero-shot)                            & 39.30 & 37.11 & 37.05 & 36.82 \\
Llama-3.3-70B + TGR + AUP + CogRAG                   & \textbf{58.23} & \textbf{57.43} & \textbf{57.48} & \textbf{57.44} \\
\bottomrule
\end{tabular}
\end{table}

\subsection{Task Design Reasons} \label{taskdesign}
The three tasks in CogLoad-Bench were selected to span distinct cognitive and sensorimotor demands~\cite{wickens2008multiple} rather than to exhaustively cover all daily activities. The Reading task engages sustained visual-linguistic processing, the Social Game task requires dynamic social cognition with visuospatial reasoning, and the Audio task involves sustained auditory attention with phonological working memory. Together, the three tasks cover different perceptual modalities (visual vs.\ auditory), processing codes (verbal vs.\ spatial), and cognitive demand types.

Beyond cognitive coverage, our data collection included realistic conditions absent in prior lab-based cognitive load studies. For example, the Social Game task involved real human partners, unpredictable verbal cues, lighting changes, and background noise, simulating the variability of real-world environments. We acknowledge that more complex tasks are necessary for fully evaluating practical generalization, and we plan to expand CogLoad-Bench with more complex scenarios such as driving and cooking in future work.

\subsection{Cross-Task Transfer and Generalization} \label{A_crosstask}
We clarify that ``cross-task generalization'' is not about exhaustive task coverage, but about the framework's ability to adapt to new tasks without model retraining (see also Sec.~\ref{taskdesign}). Unlike supervised baselines that must be retrained per task, GazeMind adapts to a new task by simply generating a new task-specific rule via TGR (Sec.~\ref{TGR}), while the LLM backbone, user profiling (AUP), and retrieval (CogRAG) all remain unchanged. Importantly, TGR rules are pre-generated from training data only, and no test samples are seen during rule generation, so evaluation is performed directly on unseen test data. The consistent performance across our evaluation tasks (Table~\ref{tab:task_results}, Figure~\ref{fig:task_analysis}) with the same unified framework validates this adaptability.

To further evaluate this claim, we test bidirectional cross-task rule transfer between the two main evaluation tasks: (1) directly applying the TGR rules generated for the Reading task to Gaming test samples, and (2) directly applying the TGR rules generated for the Gaming task to Reading test samples, both without generating any task-specific rules for the target task. Compared to using Gaming-specific rules (60.63\% accuracy, Table~\ref{tab:task_results}), direct Reading-to-Gaming rule transfer yields a 5.5 percentage point accuracy drop, but still substantially exceeds all supervised and zero-shot baselines reported in Table~\ref{tab:main_results}. In the reverse direction, Gaming-to-Reading rule transfer achieves 53.23\% accuracy, well above the best zero-shot LLM baseline (Table~\ref{tab:main_results}), despite the absence of any Reading-specific TGR rules.

This bidirectional evaluation demonstrates that even when no task-specific rules are available, GazeMind retains reasonable transferability through its other modules (TGE, AUP, CogRAG). When labeled samples for a new task are available, generating a new task-specific rule via TGR is a lightweight adaptation that does not require any model fine-tuning, fundamentally lighter than retraining supervised methods.

\subsection{Ablation Study on Temporal Window Length} \label{A_ablationtemp}

The temporal window length $T$ controls how much past gaze history is encoded in the feature table for LLM reasoning. We conduct an ablation study on $T \in \{3, 5, 10, 20\}$ seconds to examine its effect on both accuracy and inference latency. Results are summarized in Table~\ref{tab:temporal_window}.

\begin{table}[h]
\centering
\small
\caption{Ablation study on temporal window length $T$. We report accuracy and inference latency for $T \in \{3, 5, 10, 20\}$ seconds.}
\label{tab:temporal_window}
\begin{tabular}{ccc}
\toprule
\textbf{$T$ (s)} & \textbf{Accuracy (\%) $\uparrow$} & \textbf{Latency (s) $\downarrow$} \\
\midrule
3   & 51.94 & \textbf{2.13} \\
5   & \textbf{62.73} & 2.15 \\
10  & 56.14 & 2.20 \\
20  & 51.47 & 2.64 \\
\bottomrule
\end{tabular}
\end{table}

$T = 5$ achieves the best accuracy while maintaining low latency. Shorter windows ($T = 3$) lack sufficient temporal context for reliable trend estimation in the feature table, while longer windows ($T = 10$ and $T = 20$) degrade accuracy as they include noise from earlier time points that are less relevant to the current cognitive load. Latency remains relatively stable as $T$ increases, since the feature table only adds a few extra columns per additional second, contributing limited additional tokens to the LLM input.

\subsection{Task-wise Performance Analysis}\label{A_task}

\begin{figure}[h]
\centering
\includegraphics[width=\textwidth]{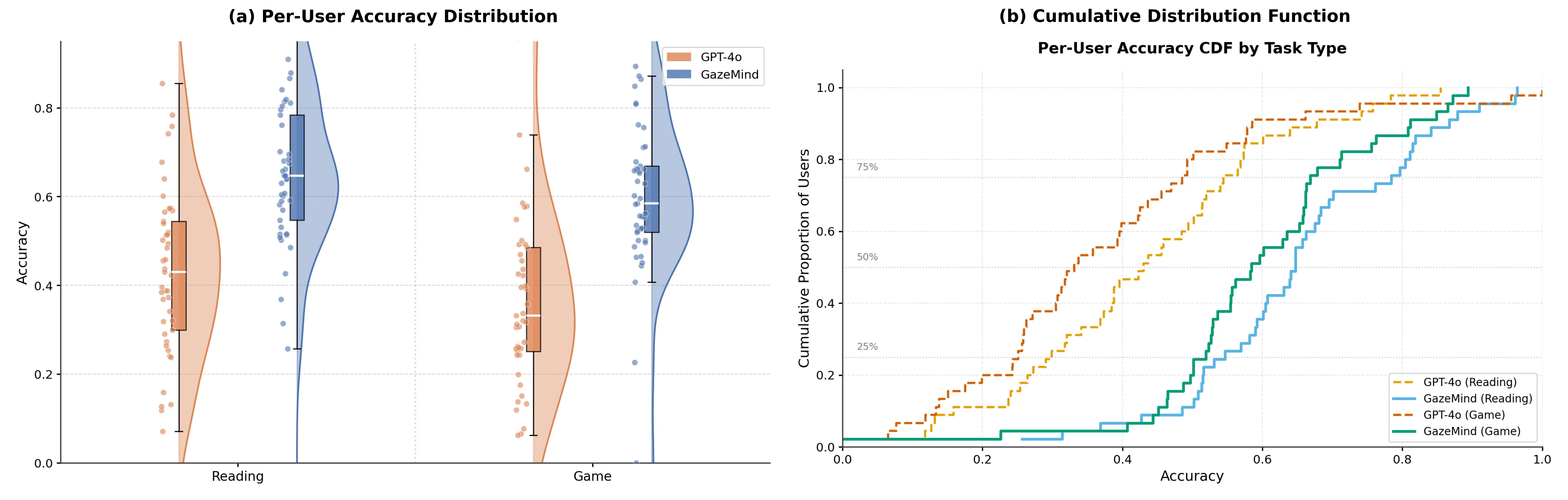}
\caption{Task-wise performance analysis comparing GazeMind and GPT-4o. (a) Per-user accuracy distributions shown via violin plots with overlaid box plots and individual data points. GazeMind (blue) consistently outperforms GPT-4o (orange) on both Reading and Game tasks, with higher medians and tighter distributions. (b) Cumulative distribution functions (CDFs) of per-user accuracy by task type. Solid lines represent GazeMind; dashed lines represent GPT-4o. The consistent rightward shift of GazeMind curves demonstrates improved accuracy across the entire user population.}
\label{fig:task_analysis}
\end{figure}

To further understand how GazeMind generalizes across different task contexts, we present a detailed task-wise analysis of per-user accuracy distributions in Figure~\ref{fig:task_analysis}.

Several observations emerge from this analysis. First, GazeMind achieves substantially higher median accuracy on both tasks compared to GPT-4o, with the improvement being consistent across the entire distribution rather than driven by outliers. Second, GazeMind exhibits lower variance in per-user accuracy, indicating more reliable performance across different individuals. Third, the Reading task yields slightly higher accuracy than the Game task for both methods, which aligns with our hypothesis that reading involves more stable and predictable gaze patterns compared to the dynamic, distraction-prone social game environment.

The CDF analysis in Figure~\ref{fig:task_analysis}(b) provides a complementary view of performance across the user population. The rightward shift of GazeMind curves relative to GPT-4o curves is evident for both tasks, indicating systematic improvement across all percentiles of users rather than gains concentrated among specific user subgroups. For GPT-4o, approximately 50\% of users achieve below 40\% accuracy on both tasks, whereas for GazeMind, the median accuracy exceeds 60\% on Reading and approaches 60\% on Game. The lower tail is also substantially lifted: fewer than 5\% of users fall below 40\% accuracy with GazeMind, compared to nearly 50\% with GPT-4o. This demonstrates that the Task-Guidance Reasoning (TGR) module effectively provides task-specific interpretation rules that enable consistent performance improvements regardless of task complexity or environmental factors.


\section{Implementation Details} \label{A}

\subsection{Implementation Details for Comparison Methods} \label{A_implementation}
All neural network baselines were trained on a single NVIDIA A100 GPU. All baselines receive the same past 5-second sliding window input and predict three-class cognitive load levels. All neural network baselines are optimized using weighted cross-entropy loss to handle class imbalance. (1) Decision Tree uses scikit-learn's implementation with Gini impurity, maximum depth of 15, minimum samples split of 50, and minimum samples leaf of 20. 
(2) Support Vector Machine (SVM) employs an RBF kernel with C=1.0 and automatic gamma scaling.
(3) Multi-Layer Perceptron (MLP) consists of three hidden layers (256-128-64 units) with ReLU activation.
(4) Long Short-Term Memory (LSTM) contains 3 hidden layers with 64 units.
(5) Temporal Convolutional Network (TCN) employs 3 temporal blocks with 64 channels, kernel size 3, and dilation rates (1, 2, 4). Each block contains two dilated causal convolutions with batch normalization, ReLU activation, dropout, and residual connections. Global average pooling aggregates temporal features before classification.
(6) Transformer uses a 2-layer Transformer encoder with hidden dimension 64 and 4 attention heads.
(7) Time-LLM adapts the time-series forecasting framework from~\cite{jin2023time} for classification. The architecture consists of a patch embedding layer projecting each timestep to LLaMA-3.1-8B's hidden dimension, a reprogramming layer with 32 learnable text prototypes using cross-attention (8 heads), and a frozen LLaMA-3.1-8B-Instruct backbone with 4-bit quantization. Only the patch embedding, positional encoding, reprogramming layer, and classification head are trainable ($\sim$ 134M parameters), while the LLaMA backbone ($\sim$ 4B parameters) remains frozen. Training uses mixed precision.

\subsection{Raw Gaze Processing \& Gaze Feature Extraction} \label{A_features}
In this section, we introduce our raw gaze processing and the gaze feature we used to represent the gaze pattern. 

The raw gaze signals are pre-processed through three steps before feature extraction. (1) The Project Aria glasses perform built-in gaze calibration before recording, ensuring baseline gaze quality. (2) Missing values and NaN samples in the raw signal are dropped. (3) Aggregating 90 raw gaze samples into one-second feature windows (90 Hz sampling rate) provides additional temporal smoothing that reduces residual noise.

Then, we extract 7 gaze features for each one-second feature window, including fixation duration, saccade duration, saccade amplitude, fixation ratio, saccade ratio, blink count, and pupil size, from the raw gaze signals. Their definitions are as follows:
\begin{itemize}
    \item \textbf{fixation duration}: The time that the eye remains relatively stationary on a specific point of interest. 
    \item \textbf{saccade duration}: The time of rapid eye movements between consecutive fixations.
    \item \textbf{saccade amplitude}: The amplitude (distance) covered by saccadic eye movements.
    \item \textbf{fixation ratio}: The proportion of time spent in focusing.
    \item \textbf{saccade ratio}: The proportion of time spent in moving.
    \item \textbf{blink count}: The frequency of eye blinks.
    \item \textbf{pupil size}: The diameter of pupil.
\end{itemize}

We employ the Velocity-Threshold Identification (I-VT) algorithm \cite{salvucci2000identifying} to classify raw gaze samples as fixations and saccades. Based on the classified gaze events, we compute five features within each one-second window. For fixation-related features, we calculate \textit{fixation duration} as the average duration across all detected fixation events, and \textit{fixation ratio} as the proportion of total window time spent in fixations. Similarly, for saccade-related features, we calculate \textit{saccade duration} as the average duration across all detected saccade events, \textit{saccade amplitude} as the average angular distance traveled from start to end point across all saccades, and \textit{saccade ratio} as the proportion of total window time spent in saccades.

\subsection{User Profiling and Clustering}\label{A_user}

To characterize individual differences in gaze behavior, we perform user profiling based on four features aggregated across all samples for each user in training set: blink intensity (mean blink count per second), pupil sensitivity (standard deviation of pupil size), pupil size baseline (mean pupil size), and gaze instability (spatial dispersion of gaze points). We apply K-Means clustering \cite{hartigan1979algorithm} on these four features to group users with similar gaze characteristics. To determine the optimal number of clusters, we evaluate Silhouette scores across different values of $K$ and find that $K=3$ yields the best separation.

\begin{figure}[h]
\centering
\includegraphics[width=0.5\textwidth]{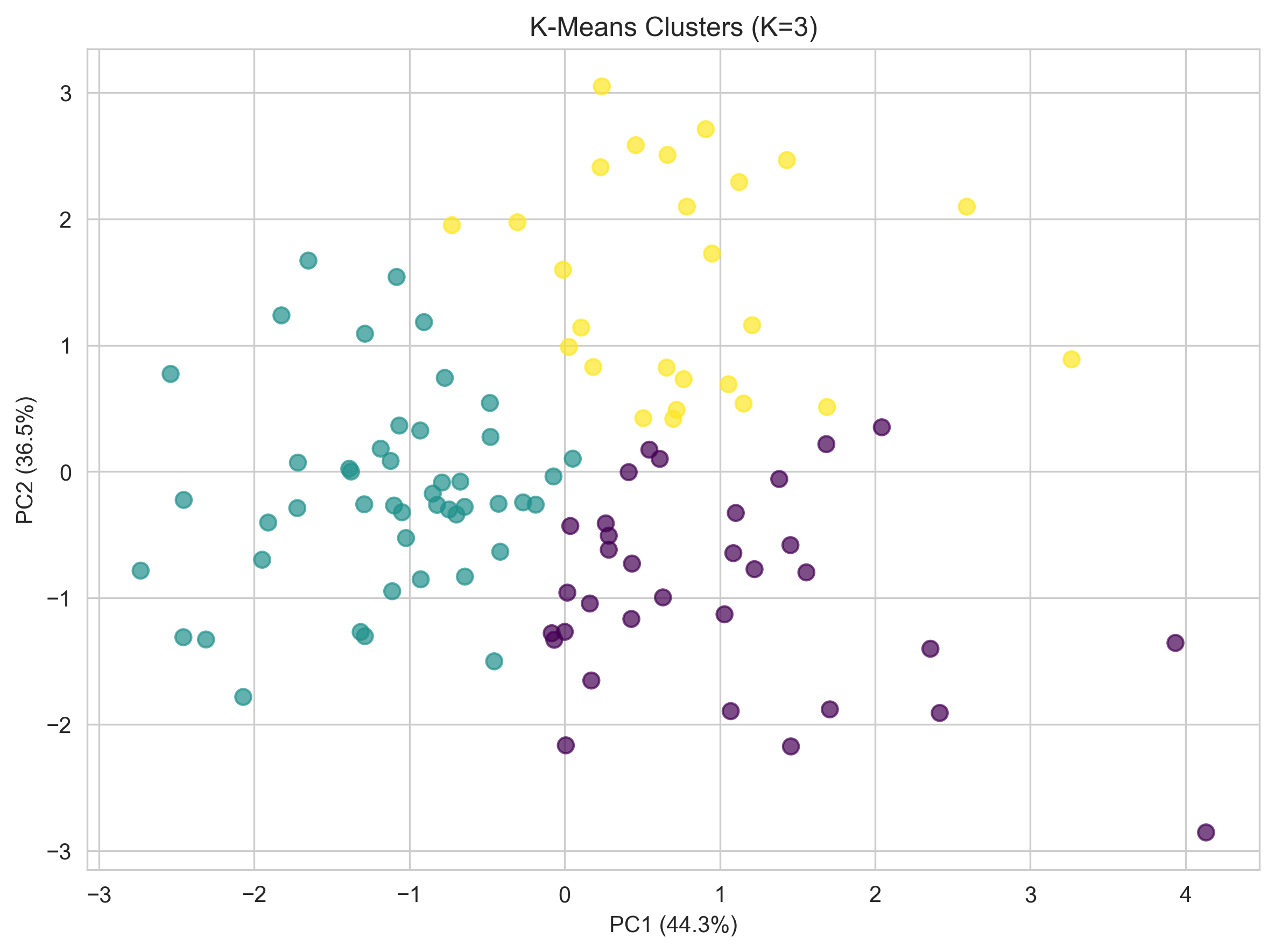}
\caption{PCA visualization of user clustering. K-Means clusters (K=3) showing three distinct user groups.}
\label{fig:user_clustering}
\end{figure}

Figure~\ref{fig:user_clustering} visualizes the clustering results using principal components analysis (PCA) \cite{dunteman1989principal} projection. It shows the three identified clusters. Based on the cluster centroids and feature distributions, we identify three distinct user profiles:

\begin{itemize}
    \item \textbf{High-Reactors}: Users with large pupil size baseline (mean=388.62) and high pupil sensitivity (std=95.31), indicating strong physiological responsiveness to cognitive demands. Blink rate and gaze instability are within normal ranges.
    
    \item \textbf{Low-Reactors}: Users exhibiting the lowest values across pupil sensitivity (std=43.86), blink rate (0.85), and gaze instability (dispersion=6.88). These users show minimal physiological variation, requiring calibrated interpretation to detect subtle cognitive load changes.
    
    \item \textbf{Restless}: Users characterized by blink rates approximately twice that of other groups (2.04) and the highest gaze instability (dispersion=11.43), combined with the smallest pupil size baseline (mean=198.87). This profile reflects naturally restless gaze behavior unrelated to cognitive load.
\end{itemize}

These profiles are incorporated into the AUP module (Sec. \ref{AUP}), enabling GazeMind to adjust its interpretation of gaze patterns based on each user's characteristic response style.

\subsection{Prompt Design}\label{A_prompt}
This section presents the complete prompt architecture used in GazeMind. The prompt consists of a \textbf{System Prompt} that defines the LLM's role and reasoning framework, and a \textbf{User Prompt} that provides session-specific context and data.

The system prompt contains all task-agnostic components that remain constant across inference sessions, including the LLM's role definition, feature interpretation guidelines, feature definitions, user profile understanding, guidance of the use of referenced examples, and the four-phase internal reasoning process framework. The user prompt, in contrast, contains only session-specific information: task guidance rule generated by TGR, user profile and traits from AUP, retrieved reference examples from CogRAG, and the current feature table from TGE.

The four-phase reasoning framework operationalizes the unified inference described in Sec. \ref{unified} by providing explicit context guidance to the LLM. Phase 1 instructs the model to first establish an understanding of the current user by parsing their assigned profile type, associated traits, and personalized gaze baselines (pupil size, blink count) computed from calibration data. This user-specific context enables the LLM to adjust its interpretation of gaze patterns accordingly throughout the subsequent phases. Phase 2 combines task-specific rules from TGR with user profile adjustment from AUP to form a provisional classification. Phase 3 implements the reference data verification protocol, requiring the LLM to qualitatively validate retrieved examples by comparing dominant physiological states before using them, and to fall back to calibrated task logic if validation fails. Phase 4 performs consistency checks across all evidence sources and resolves conflicts using a clear priority hierarchy: validated reference examples take precedence over calibrated task logic, which in turn takes precedence over raw feature interpretation. This prioritization reflects the principle that validated empirical ground-truth from similar users provides more reliable guidance than population-derived rules for edge cases.

\subsubsection{System Prompt}
The complete system prompt is shown below:

\begin{tcolorbox}[colback=gray!5!white, colframe=gray!75!black, title=System Prompt, breakable, fonttitle=\bfseries\small, fontupper=\scriptsize]
\begin{verbatim}
## Role
You are an expert in Neuroergonomics and cognitive load assessment using eye-tracking data. 
Your objective is to classify a user's Cognitive Load Level as 'Low', 'Moderate', or 'High' based on 
a {time_window}-second window of gaze features.

## Data Interpretation Guide 
CRITICAL: The input data are Z-Scores (Standard Deviations from the GLOBAL POPULATION baseline).
You must interpret the Magnitude and Direction separately:

1. Magnitude:
   * Near 0.0 (-0.5 to +0.5): Baseline behavior. Normal state.
   * Significant Deviation (> +1.0 or < -1.0): A strong physiological signal.
   * Extreme Deviation (> +2.5 or < -2.5): Very intense event.

2. Direction:
   * Positive (+): Value is Higher/Longer/More Frequent than usual.
   * Negative (-): Value is Lower/Shorter/Suppressed compared to usual.
   * WARNING: A Negative Z-Score does NOT automatically mean "Low Load". In some tasks, "Shorter"
     fixations (Negative Z) often indicate High Load.

## Feature Definitions
* fix_dur: Duration of fixations. (Pos=Long Stare; Neg=Rapid Glances)
* sac_dur: Duration of saccades.
* sac_amp: Amplitude of eye movement. (Pos=Wide Scanning; Neg=Tunnel Vision)
* fix_ratio: Proportion of time spent focusing.
* sac_ratio: Proportion of time spent moving.
* blink_count: Frequency of blinks.
* avg_pupil_size: Average pupil diameter.

## Calibration Protocol (User-Adaptive Logic)
You will be provided with specific Task Logic (Standard Rules) and a User Profile & Calibration
Instructions (Calibration).
1. Standard Rules are concluded based on the global population.
2. User Profile & Calibration Instructions give guidance to adjust some specific guidance in the 
Standard Rules.

## Reference Data Verification Protocol
You will receive Reference Data containing REAL historical examples from similar users. These 
examples are retrieved based on mathematical similarity to the current session feature. You must 
qualitatively validate retrieved examples before using them. Mathematical similarity does not 
guarantee physiological relevance.

## Internal Reasoning Process
Follow this four-phase reasoning process:

Phase 1: Understand User Profile
- Understand the avg_pupil_size and blink_count baselines from the User Profile. Identify the user 
profile type and traits.

Phase 2: Task Logic Application & Calibration
- Step A: Apply the standard rules from the Task Logic section to the Feature Data.
- Step B: ADJUST interpretation based on User Profile & Calibration Instructions.
- Step C: Form a provisional classification based on Task Logic + User Profile.

Phase 3: Reference Data Verification
- Step A (Qualitative Validation): Compare the dominant physiological state (Active, 
Suppressed, or Stable) of Reference Examples against the Current Session. DISCARD any example showing 
a conflicting pattern.
- Step B (Evidence Prioritization): If a validated Reference Example contradicts your Provisional 
Classification, TRUST THE REFERENCE EXAMPLE.
- Step C (Fallback): If Reference Examples fail validation, revert to the Calibrated Task Logic from 
Phase 2.

Phase 4: Final Consistency Reflection
- Cross-validate: Does the conclusion align with User Profile traits? Does it satisfy calibrated
  Task Logic? Is it consistent with validated Reference Examples?
- Conflict Resolution: Prioritize evidence in this order: Validated Reference Examples > Calibrated
  Task Logic > Raw Feature Data.

## Output Requirements
Cognitive Load: [Low/Moderate/High]
Reasoning: brief explanation in few sentences, summarizing how you arrived at the conclusion.
\end{verbatim}
\end{tcolorbox}

\subsubsection{User Prompt}

The user prompt provides session-specific information including task context, user profile, retrieved reference examples, and the current gaze feature data. The complete user prompt template is shown below:

\begin{tcolorbox}[colback=gray!5!white, colframe=gray!75!black, title=User Prompt, breakable, fonttitle=\bfseries\small, fontupper=\scriptsize]
\begin{verbatim}
## Current Session
Task: {task_name}
Time Window: {time_window} seconds

## Task Logic
{task_guidance}

## User Profile & Calibration Instructions
{user_profile_traits}

## Reference Data (Retrieval-Augmented Context)
{rag_context}

## Feature Data (Z-Scores)
{feature_table}

## Request
Based on the Task Logic (Standard), User Profile & Calibration Instructions (Adjust), AND Reference 
Data (Ground Truth Examples) above, classify the cognitive load.

## Output Requirements
Cognitive Load: [Low/Moderate/High]
Reasoning: brief explanation in few sentences, summarizing how you arrived at the conclusion.
\end{verbatim}
\end{tcolorbox}

\section{Cogload-Bench Dataset Details} \label{A_data}

This section provides additional details on the CogLoad-Bench dataset, including participant demographics, detailed task protocols, annotation procedures, and label distributions.



\subsection{Participant Demographics}

Eligibility criteria included fluency in English, no self-reported vision disorders, and no drug or alcohol use within 24 hours of participation. A total of 152 participants were recruited for this study.

\subsubsection{Age and Sex Distribution}

\begin{figure}[h]
\centering
\includegraphics[width=0.9\textwidth]{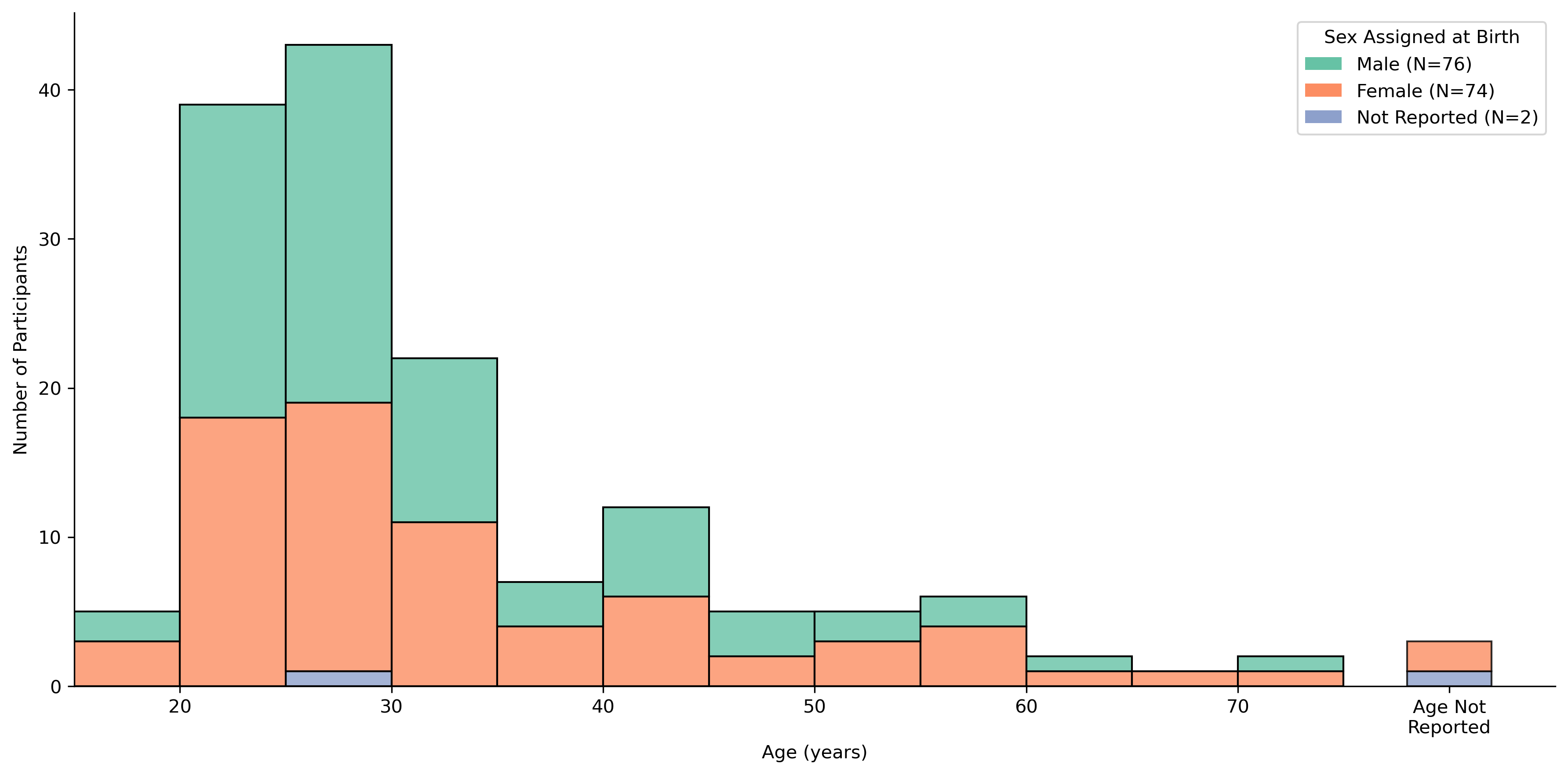}
\caption{Age and sex distribution of CogLoad-Bench participants. The histogram shows participant counts across age bins, with colors indicating sex assigned at birth. The sample is approximately balanced by sex (Male: N=76, Female: N=74, Not Reported: N=2) and spans ages 18--72 years.}
\label{fig:age_sex}
\end{figure}

Participants ranged in age from 18 to 72 years (M = 31.9, SD = 11.58). The sample was approximately balanced by sex assigned at birth, with 76 males (50.0\%), 74 females (48.7\%), and 2 participants who did not report (1.3\%). As shown in Figure~\ref{fig:age_sex}, the majority of participants were between 20 and 35 years old. The sex distribution remained relatively balanced across age groups.

\subsubsection{Racial and Ethnic Identity}

\begin{figure}[h]
\centering
\includegraphics[width=0.7\textwidth]{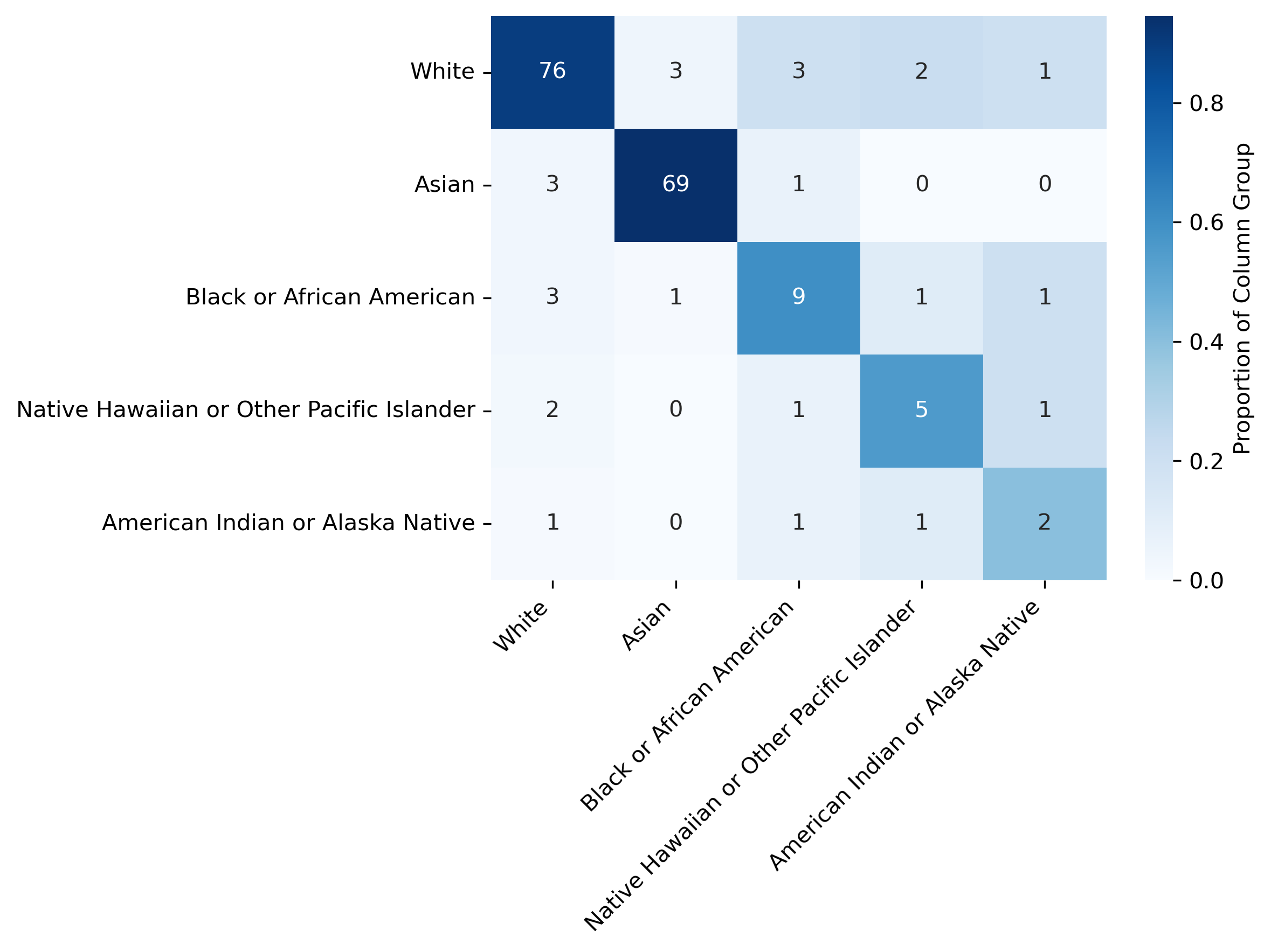}
\caption{Racial identity co-occurrence matrix. Numbers indicate participant counts. Diagonal entries represent participants identifying with a single racial category; off-diagonal entries represent participants selecting multiple racial identities. Color intensity reflects the proportion within each column group.}
\label{fig:racial}
\end{figure}

Participants self-reported their racial identity, with the option to select multiple categories. The sample consisted primarily of participants identifying as White (N=76) or Asian (N=69), with smaller proportions identifying as Black or African American (N=9), Native Hawaiian or Other Pacific Islander (N=5), or American Indian or Alaska Native (N=2). Figure~\ref{fig:racial} presents the co-occurrence matrix of racial identities, where diagonal entries indicate participants who selected only that category, and off-diagonal entries show participants who selected multiple racial identities. The matrix reveals that most participants identified with a single racial category, while a small subset reported multiple racial identities.

Additionally, the sample was predominantly right-handed (90.5\%).

\subsection{Detailed Task Protocols}

\subsubsection{Audio Task}

Participants completed an auditory N-Back task, a standard paradigm for manipulating cognitive load, in which they responded verbally when the current letter matched a letter presented a specified number of positions earlier in the sequence (1-, 2-, or 3-back), depending on the block. Blocks were presented in a fixed order (1-back $\rightarrow$ 2-back $\rightarrow$ 3-back $\rightarrow$ 2-back $\rightarrow$ 1-back $\rightarrow$ 2-back $\rightarrow$ 3-back $\rightarrow$ 2-back $\rightarrow$ 1-back) to induce systematic increases and decreases in cognitive load. Each block lasted 30 seconds, with letters presented auditorily every 2.5 seconds. Participants reported their cognitive load approximately 15 seconds into each block and again at the end of the block.

\subsubsection{Reading Task}

Participants completed a reading-based reasoning task in which task difficulty was manipulated through question design. Questions belonged to two varieties: \emph{Days of the Week} problems, which required reasoning about relative temporal relationships (e.g., ``What day is two days after the day before Thursday?''), and \emph{Transaction} problems, involving mental arithmetic (e.g., ``If you have \$20 and spend \$7, then receive \$3...'').

For both varieties, questions were organized into three difficulty levels (easy, medium, hard), defined by the amount of information and complexity of reasoning operations needed to solve the problem. Questions were drawn randomly from a pre-generated pool and presented for approximately one minute per difficulty level. Starting difficulty was randomized, with subsequent levels following a fixed progression (e.g., easy $\rightarrow$ medium $\rightarrow$ hard $\rightarrow$ medium $\rightarrow$ easy). Participants silently read each question on a laptop screen and responded verbally without external aids (e.g., pens, paper, or finger counting). Participants reported their cognitive load immediately after responding to each question and before the correct answer was revealed.

\subsubsection{Social Game Task}

Participants completed an interactive social game based on a modified version of \emph{Taco Cat Goat Cheese Pizza}. Task difficulty was manipulated across three levels (easy, medium, hard) by varying gameplay structure and distraction:

\begin{itemize}
    \item \textbf{Easy:} Gameplay followed a fixed sequence with no competing players.
    \item \textbf{Medium:} Participants alternated turns with the experimenter.
    \item \textbf{Hard:} Difficulty was further increased via unpredictable verbal cues or background auditory distraction.
\end{itemize}

Each difficulty level lasted approximately one minute, with starting difficulty randomized and subsequent levels following a fixed progression as in the reading task. Participants reported their cognitive load roughly every 30 seconds without interrupting gameplay.

\subsection{Cognitive Load Annotation}

\subsubsection{Definition}

In this study, we focus on the mental effort component of cognitive load, defined as ``the cognitive capacity that is actually allocated to accommodate the demands imposed by the task''~\cite{paas1992training}. This definition is particularly relevant in working and learning contexts and aligns with established subjective measurement approaches in cognitive load research.

\subsubsection{Real-Time Self-Report Protocol}

Cognitive load annotations were obtained via frequent (roughly every 15 to 30 seconds) self-reports collected during task execution. Participants responded verbally to an auditory cue by reporting their perceived cognitive load on a 7-level ordinal scale: \emph{very low}, \emph{low}, \emph{moderate-low}, \emph{moderate}, \emph{moderate-high}, \emph{high}, and \emph{very high}. A structured pretest ensured consistent understanding and use of the scale across participants.

Frequent in-task reporting is a key strength of this dataset, as subjective measures of mental effort are most valid when collected during task performance rather than retrospectively~\cite{van2012timing}.

Verbal cognitive load reports were transcribed using an automatic speech-to-text pipeline based on the Whisper model~\cite{radford2023robust}. A representative sample of transcripts were manually reviewed to ensure valid label extraction.

\subsubsection{Label Aggregation}

For modeling, the 7-level annotations were collapsed into a 3-level label representation to improve robustness and reduce ambiguity in ordinal boundaries:

\begin{itemize}
    \item \textbf{Low:} very low, low, moderate-low
    \item \textbf{Moderate:} moderate
    \item \textbf{High:} moderate-high, high, very high
\end{itemize}

This aggregation mitigates inter-participant variability in scale usage while preserving meaningful distinctions in cognitive load.
\begin{figure}[t]
\centering
\includegraphics[width=\textwidth]{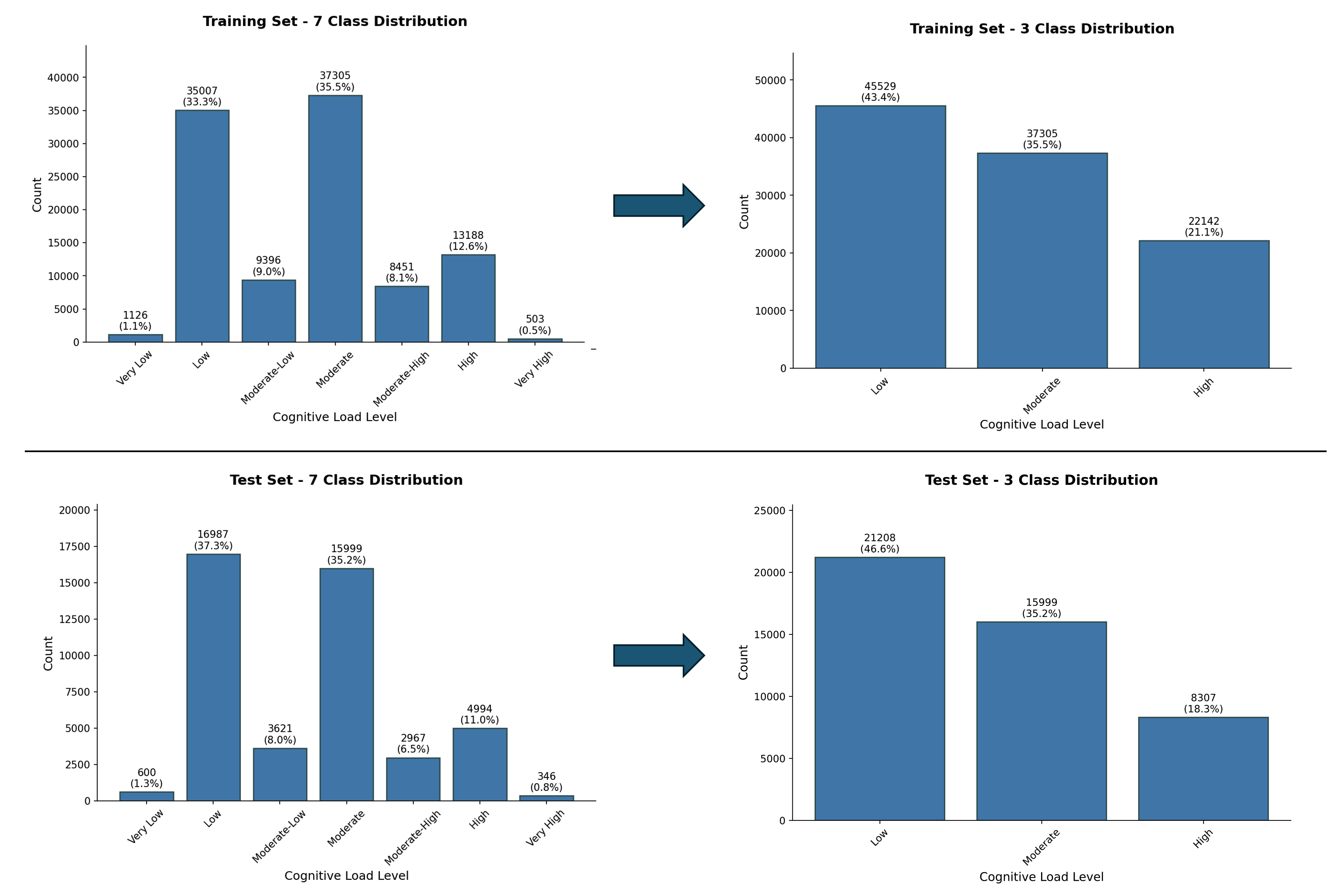}
\caption{Cognitive load label distributions in CogLoad-Bench. \textbf{Left column:} Original 7-class distributions. \textbf{Right column:} Aggregated 3-class distributions used for modeling. \textbf{Top row:} Training set (106 participants). \textbf{Bottom row:} Test set (46 participants). The arrows indicate the label aggregation process from 7 classes to 3 classes.}
\label{fig:label_dist}
\end{figure}
\subsubsection{Temporal Label Interpolation}

Annotations were collected at discrete time points and interpolated within each task to produce time-aligned cognitive load labels suitable for time-series modeling. Interpolation was performed within task boundaries only, excluding instruction and other non-task periods, and followed task-specific rules reflecting differences in task structure and query timing:

\begin{itemize}
    \item \textbf{Audio N-Back:} Each annotation was propagated backward in time to the start of the most recent continuous letter stream, excluding instruction periods without auditory stimuli.
    \item \textbf{Reading Task:} Annotations were propagated backward to the preceding load report, which occurred after the participant answered each question.
    \item \textbf{Social Game:} Annotations were propagated asymmetrically, with 80\% of the interval assigned backward to the preceding report and 20\% assigned forward to the subsequent report.
\end{itemize}

This procedure yields piecewise-constant, temporally aligned cognitive load labels that capture within-task dynamics and are well suited for supervised learning on multimodal time-series data.

\subsection{Label Distribution}

Figure~\ref{fig:label_dist} presents the cognitive load label distributions for both training and test sets, showing both the original 7-class annotations and the aggregated 3-class labels used for modeling.

In the original 7-class annotation, both training and test sets show a concentration of labels around \emph{Low} and \emph{Moderate} levels, with relatively fewer samples at the extreme ends (\emph{Very Low} and \emph{Very High}). This distribution reflects typical self-reporting behavior where participants tend to avoid extreme ratings. After aggregation to 3 classes, the distribution becomes more balanced. The similar distributions between training and test sets indicate successful stratification despite the cross-user split, ensuring fair evaluation of model generalization.

\subsection{Dataset Statistics Summary}

Table~\ref{tab:dataset_stats} summarizes the key statistics of CogLoad-Bench.

\begin{table}[h]
\centering
\caption{CogLoad-Bench dataset statistics.}
\label{tab:dataset_stats}
\resizebox{0.5\linewidth}{!}{
\begin{tabular}{lc}
\toprule
\textbf{Statistic} & \textbf{Value} \\
\midrule
Total participants & 152 \\
Training set participants & 106 (70\%) \\
Test set participants & 46 (30\%) \\
Total recordings & 456 \\
Total duration & 40+ hours \\
Duration per participant & $\sim$15 minutes \\
Duration per task & $\sim$5 minutes \\
\midrule
Total annotations & 150490 \\
\quad Training set & 104976 \\
\quad Test set & 45514 \\
\midrule
Eye tracking frequency & 90 Hz \\
Video frequency & 10 Hz \\
Audio frequency & 48 kHz \\
\midrule
Age range & 18--72 years \\
Mean age (SD) & 31.9 (11.58) \\
Sex (Male / Female / Not Reported) & 76 / 74 / 2 \\
\bottomrule
\end{tabular}}
\end{table}

\section{Impact Statement} \label{A_privacy}

GazeMind enables proactive, context-aware AI assistants on smart glasses by inferring users' cognitive load from eye gaze. This capability has potential to improve user experience by reducing interruptions during high-demand tasks and supporting accessibility applications such as adaptive learning tools or assistive interfaces for individuals with cognitive challenges. At the same time, the underlying technology raises important ethical and privacy concerns that we address from four perspectives.

\paragraph{Privacy.}
GazeMind's retrieval operates on statistical descriptors of gaze patterns rather than raw gaze signals, and the framework supports on-device inference (Table~\ref{tab:small_llm}), avoiding transmission of sensitive data to remote servers. Prior work on privacy-preserving gaze streaming~\cite{wilson2024privacy} and lightweight gaze privacy techniques~\cite{raju2025real} is directly applicable to our pipeline.

\paragraph{Anonymization.}
User profiles are stored as cluster types (e.g., ``Low-Reactor'') without any personally identifiable information.

\paragraph{Misclassification safeguards.}
Each prediction is accompanied by an explanation that supports human verification, and conflicting evidence is flagged rather than blindly trusted (e.g., Figure~\ref{case_figure}, Case 2), reducing the risk of unchecked errors propagating into downstream actions.

\paragraph{Misuse prevention.}
We recognize the risk that cognitive load inference could be misused for surveillance, such as workplace monitoring of worker attention or productivity. Responsible deployment must require explicit user consent and opt-out, and cognitive load data should be accessible only to the user, not to third parties.


\end{document}